# Do stars still form in molecular gas within CO-dark dwarf galaxies?

David J. Whitworth,[1,2,3]* Rowan J. Smith,[4,2] Simon C. O. Glover,[3] Robin Tress,[5] Elizabeth J Watkins,[2] Jian-Cheng Feng,[2,6,7] Noé Brucy,[3,8] Ralf S. Klessen,[3,9,10,11] Paul C. Clark[12]

[1] *Universidad Nacional Autónoma de México, Instituto de Radioastronomía y Astrofísica, Antigua Carretera a Pátzcuaro 8701, Ex-Hda. San José de la Huerta, 58089 Morelia, Michoacán, México*
[2] *Jodrell Bank Centre for Astrophysics, Department of Physics and Astronomy, University of Manchester, Oxford Road, Manchester M13 9PL, UK*
[3] *Universität Heidelberg, Zentrum fur Astronomie, Institut fur Theoretische Astrophysik, Albert-Ueberle-Str. 2, 69120 Heidelberg, Germany*
[4] *SUPA School of Physics and Astronomy, University of St Andrews, North Haugh, St Andrews, Fife, KY17 9SS, UK*
[5] *Institute of Physics, Laboratory for Galaxy Evolution and Spectral Modelling, EPFL, Observatoire de Sauverny, Chemin Pegasi 51, 1290 Versoix*
[6] *Purple Mountain Observatory, Chinese Academy of Sciences, No.10 Yuanhua Road, Qixia District, Nanjing 210023, China*
[7] *University of Science and Technology of China, No.96, JinZhai Road Baohe District, Hefei, Anhui, 230026, China*
[8] *Centre de Recherche Astrophysique de Lyon UMR5574, ENS de Lyon, Univ. Lyon1, CNRS, Université de Lyon, 69007, Lyon, France*
[9] *Universität Heidelberg, Interdisziplinäres Zentrum für Wissenschaftliches Rechnen, Im Neuenheimer Feld 205, D-69120 Heidelberg, Germany*
[10] *Harvard-Smithsonian Center for Astrophysics, 60 Garden Street, Cambridge, MA 02138, USA*
[11] *Elizabeth S. and Richard M. Cashin Fellow at the Radcliffe Institute for Advanced Studies at Harvard University, 10 Garden Street, Cambridge,*
[12] *School of Physics and Astronomy, Queens Buildings, The Parade, Cardiff University, Cardiff, CF24 3AA*

**ABSTRACT**
In the Milky Way and other main-sequence galaxies, stars form exclusively in molecular gas, which is traced by CO emission. However, low metallicity dwarf galaxies are often 'CO-dark' in the sense that CO emission is not observable even at the high resolution and sensitivities of modern observing facilities. In this work we use ultra high-resolution simulations of four low-metalicity dwarf galaxies (which resolve star formation down to the scale of star-forming cores, 0.01 pc) combined with a time-dependent treatment of the chemistry of the interstellar medium, to investigate the star formation environment in this previously hidden regime. By generating synthetic observations of our models we show that the galaxies have high to extremely high dark gas fractions (0.13 to 1.00 dependent on beam size and conditions), yet despite this form stars. However, when examined on smaller scales, we find that the stars still form in regions dominated by molecular gas, it is simply that these are far smaller than the scale of the beam ($1.5''$). Thus, while stars in CO-dark dwarf galaxies form in small molecular cores like larger galaxies, their cloud-scale environment is very different.

## 1 INTRODUCTION

Studies of tracers of molecular gas in the Milky Way and nearby spiral galaxies have shown that there is a strong relationship between star formation and molecular gas (Kennicutt 1998; Leroy et al. 2008; Krumholz & McKee 2005; McKee & Ostriker 2007; Ostriker et al. 2010; Bigiel et al. 2011). However, in low metallicity systems $H_2$ formation is inefficient due to the low dust content (Palla et al. 1983; Omukai et al. 2010), which prompts the obvious question of whether the relationship between molecular gas and star formation is the same in these systems as in more metal-rich spirals.

Unfortunately, it is difficult to answer this question purely observationally. Although $H_2$ is the most abundant molecule in the Universe, it does not emit efficiently at the low temperatures ($\sim 10$ K) found in the cold, dense phase of the interstellar medium (ISM) where star formation occurs. $H_2$ is a low mass symmetric molecule with no electric dipole moment, and as a result, its lowest accesible rotational transition occurs at $\sim 510$ K (Dabrowski 1984; Goldsmith et al. 2010). Collisional excitation of this transition in gas with a temperature of a few tens of Kelvin is therefore almost impossible. As a result we are forced to rely on alternative tracers of the molecular gas that do emit efficiently at low temperatures. Carbon monoxide (CO) is one such molecule due to its low excitation energy of $\sim 5$ K for the $J = 1 - 0$ ground state transition. It is often used as a tracer for $H_2$ in the ISM (see e.g. Bolatto et al. 2013, Chevance et al. 2022 and references therein).

However, the potential existence of clouds of 'CO-dark'

* E-mail: d.whitworth@irya.unam.mx

molecular gas poses a problem in our understanding of the role that molecular gas plays in metal-poor star-forming galaxies. The greater effectiveness of $H_2$ self-shielding compared to CO self-shielding allows $H_2$ to survive in the ISM at lower column and volume densities than CO. We therefore expect to find regions of molecular gas with negligible CO at the outskirts of giant molecular clouds (GMCs) (Wolfire et al. 2010; Clark et al. 2012; Glover & Clark 2016; Klessen & Glover 2016) and potentially also clouds of $H_2$ that are completely CO-dark. If molecular clouds are therefore bigger than predicted due to CO-dark gas then cloud properties and star formation derived from CO could be misleading, leading to under-estimated results.

In the Milky Way, estimates of the fraction of $H_2$ in CO-dark regions range from $f_{\rm DG} = 0.3$ to $f_{\rm DG} = 0.7$ (Wolfire et al. 2010; Smith et al. 2014; Seifried et al. 2020). Moreover, it has been suggested that there could be a metallicity scaling factor needed to convert CO luminosity to molecular mass to account for environmental conditions (Evans et al. 2022). There are also indications that the value becomes much larger in metal-poor systems. For example, Chevance et al. (2020c) find that $f_{\rm DG} > 0.75$ in the 30 Doradus star-forming region in the Large Magellanic Cloud (LMC), Tokuda et al. (2021) find a value of $f_{\rm DG} \sim 0.9$ in the SMC, and Madden et al. (2020) report even larger fractions in their sample of metal-poor dwarf galaxies. In cases where this CO-dark material is well-mixed with CO-bright molecular gas (as in Milky Way GMCs, for example), its presence may explain some of the scale-dependence observed in our estimates of $X_{\rm CO}$: measurements made on larger scales will tend to include more CO-dark $H_2$ and hence will yield a larger value for $X_{\rm CO}$ than measurements made on the scale of individual CO-bright cores (Gong et al. 2020, & references therein). However, molecular clouds that are completely undetected in CO are much harder to account for without a good model of how $f_{\rm DG}$ varies with metallicity.

Observations of star formation in dwarf galaxies that is uncorrelated with detectable CO emission (e.g. Cormier et al. 2017) may therefore indicate that stars are forming in CO-dark molecular clouds in these galaxies. However, as we expect star formation in low metallicity environments to be more strongly correlated with cold gas than with molecular gas (Glover & Clark 2012; Whitworth et al. 2022; Glover 2023), it is also possible that this star formation is occurring in clouds dominated by atomic rather than molecular gas. Further complicating efforts to distinguish between these scenarios observationally is the fact that we expect star-forming regions to rapidly decouple from their birth clouds on timescales of only a few Myr (Schruba et al. 2010; Onodera et al. 2010; Grasha et al. 2018, 2019; Kreckel et al. 2018; Kruijssen et al. 2018, 2019; Schinnerer et al. 2019; Chevance et al. 2020b), meaning that is is possible to directly observe the birth environments of only the very youngest regions.

For this reason, numerical simulations are invaluable for helping us to better understand the relationship between $H_2$, CO and star formation in metal-poor CO-dark galaxies. In this paper, we analyse results from four extremely high resolution hydrodynamical simulations of very low metallicity dwarf galaxies, in which we have independently varied the metallicity ($Z = 0.1, 0.01 Z_\odot$) and the strength of the interstellar radiation field ($G_0 = 0.1, 0.01$). We select several 100 pc by 100 pc sub-regions from within each of the models and post-process them to produce synthetic observations of the integrated velocity of $^{12}$CO J=1-0, $W_{\rm CO(1-0)}$. Using these synthetic emission maps, we determine the dark gas fraction of each model. We look at the stellar ages of all sink particles in each model and compare their ages and distribution to the density distribution of molecular gas, cold gas and CO. We also investigate the effects of observational beam size and selection bias.

The structure of our paper is as follows. In Section 2 we describe our simulations and the increase in resolution. In Section 3 we describe the CO distribution, radiative transfer model applied to obtain the synthetic observations and dark gar results. In Section 4, we present a discussion of the star-forming region in the simulations in terms of the age distribution and relation to gas density. We discuss the caveats of our study in Section 5 and provide a summary of the work in Section 6.

## 2 METHOD

### 2.1 Numerical model

#### 2.1.1 Basic parameters

The models we use in our analysis were first presented in Whitworth et al. (2022), hereafter Paper I. We modelled four metal-poor isolated dwarf galaxies (denoted hereafter as Z.10 G.10, Z.10 G.01, Z.01 G.10 and Z.01 G.01) using the AREPO moving mesh hydrodynamical code (Springel 2010), varying both the metallicity (Z) and interstellar radiation field (G); see Table 1 for details of the variables.

For the work presented in this study, we reran new versions of these simulations starting from 1 Gyr and continuing for 15 Myr to ensure we are in a steady state period of galactic evolution, with an increased resolution across the box (see section 2.1.2) and made synthetic observations of selected regions with the radiative transfer code POLARIS. Figure 1 shows the full gas distribution of our models at the point of analysis in this paper. In the remainder of this section, we present a brief overview of the simulations, but we refer the reader to Paper I for a more detailed description.

AREPO solves the hydrodynamic equations on an unstructured Voronoi mesh which allows the cells to move with the local gas velocity. After each timestep the mesh is reconstructed. This means the spatial resolution of the mesh can vary naturally with the local density of the gas, while avoiding the substantial mesh distortion. By default, AREPO refines and/or derefines the mesh to keep the mass of each mesh cell within a factor of two of a base mass resolution, which in Paper I we took to be $100\,{\rm M_\odot}$ for the first 300 Myr of the simulation and $50\,{\rm M_\odot}$ thereafter (see section 2.1.2 for a more detailed discussion on resolution). This default refinement criterion is supplemented by an additional Jeans refinement criterion: we ensure that the Jeans length is resolved by at least 8 mesh cells throughout the simulation in order that gravitationally collapsing gas is resolved and artificial fragmentation is avoided (Jeans 1902; Truelove et al. 1997; Federrath et al. 2011). In the simulations from Paper I, this yields a cell size of $\sim 0.16\,{\rm pc}$ at the sink density threshold of $10^4\,{\rm cm^{-3}}$. In the new simulations presented in this paper, the cell size at sink formation is smaller, $\sim 0.02\,{\rm pc}$, owing to our base mass now being set at $1\,{\rm M_\odot}$ and our choice of a higher density threshold for sink particle formation. Further details can be found in Figure 2 and Section 2.1.2.

| Model | Z($Z_\odot$) | Dust to-gas | $\zeta_H$ (s$^{-1}$) | $G_0$ (Habing units) |
|---|---|---|---|---|
| Z.10 G.10 | 0.10 | 0.10 | 3.0 ×10$^{-18}$ | 0.17 |
| Z.10 G.01 | 0.10 | 0.10 | 3.0 ×10$^{-19}$ | 0.017 |
| Z.01 G.10 | 0.01 | 0.01 | 3.0 ×10$^{-18}$ | 0.17 |
| Z.01 G.01 | 0.01 | 0.01 | 3.0 ×10$^{-19}$ | 0.017 |

**Table 1.** Values used for the metallicity ($Z$), dust-to-gas ratio (relative to the value in solar metallicity gas), cosmic ray ionisation rate of atomic hydrogen ($\zeta_H$) and UV field strength ($G_0$).

Our model dwarf galaxies are set up as isolated systems with a stable disc consisting of two components: a dark matter halo and a gaseous disc. A stellar population is not included in the initial set up, but instead forms naturally over time as the gas forms stars. For the dark matter halo, we use a spheroidal Hernquist (1990) profile following:

$$\rho_{\rm sph}(r) = \frac{M_{\rm sph}}{2\pi} \frac{a}{r(r+a)^3}, \quad (1)$$

where $r$ is the radius of the sphere, $a$ is the scale-length of the halo (7.62 pc), and M$_{\rm sph}$ is the mass of the halo, which we take to be $1.99 \times 10^{10}$ M$_\odot$.

The gaseous disc component in the initial models from Paper I follows a double exponential density profile:

$$\rho_{\rm disc}(R,z) = \frac{M_{\rm disc}}{2\pi h_z h_R^2} {\rm sech}^2\left(\frac{z}{2h_z}\right) \exp\left(-\frac{R}{h_R}\right), \quad (2)$$

where $R$ is the disc radius and $z$ is its height, $h_z$ and $h_R$ are the scale-height (0.35 kpc) and scale-length (0.82 kpc) of the disc, respectively and M$_{\rm disc}$ is the mass ($8.00 \times 10^7$ M$_\odot$) of the disc. The initial conditions were generated using the method from Springel (2005) and chosen to be broadly comparable to Hu et al. (2016).

The chemical and thermal evolution of the interstellar medium (ISM) is modelled using a non-equilibrium, time-dependent chemical network based on the work of Gong et al. (2017) with the modifications discussed in Hunter et al. (2023). This network allows us to directly trace the non-equilibrium chemical abundances of 9 chemical species (H$_2$, H$^+$, C$^+$, CH$_x$, OH$_x$, CO, HCO$^+$, He$^+$ and Si$^+$; note that OH$_x$ and CH$_x$ are composite species representing hydroxyl and related compounds, and light hydrocarbons (CH, CH$^+$, etc.) respectively). The abundances of a further 8 species (free electrons, H, H$_3^+$, C, O, O$^+$, He and Si) follow from conservation laws or the assumption of chemical equilibrium. Shielding from dust, as well as H$_2$ and CO self-shielding, are modelled using the TreeCol algorithm (Clark et al. 2012), with a maximum shielding length of 30 pc. Radiative heating and cooling are modelled using the atomic and molecular cooling functions from Clark et al. (2019). The values of the metallicity Z(Z$_\odot$), dust-to-gas ratio (in units of the solar neighbourhood value), cosmic ray ionisation rate, and UV field strength $G_0$ (in Habing units) that are adopted in each simulation are listed in Table 1. Initially, the gas is fully atomic with a temperature of $T = 10^4$ K.

| | |
|---|---|
| $\rho_c$ (g cm$^{-3}$) | $2.4 \times 10^{-19}$ |
| n (cm$^{-3}$) | $10^5$ |
| $r_{\rm acc}$ (pc) | 0.01 |
| Softening Length (pc) | 0.01 |
| Max sink mass (M$_\odot$) | 1000 |
| Base cell mass (M$_\odot$) | 1 |
| $\epsilon_{\rm SF}$ | 0.32 |
| $r_{\rm SNe}$ (pc) | 5 |

**Table 2.** Sink particle and resolution parameters used. $\rho_c$ is the mass density threshold for sink creation, n (cm$^{-3}$) is the corresponding number density, $r_{\rm acc}$ is the sink accretion radius, the softening length is the initial softening length for the gas cells before adaptive softening starts, the base cell mass is the target mass of the cell that can vary by a up to a factor of 2 (and may be further refined to resolve the Jeans length), $\epsilon_{\rm SF}$ is the star formation efficiency and $r_{\rm SNe}$ is the radius of scatter for SNe around the sink particle.

### 2.1.2 Resolution

Although the models presented in Paper I were carried out at high resolution, with a minimum cell size of $\sim 0.16$ pc at the density threshold for sink formation ($n = 10^4$ cm$^{-3}$), this is not sufficient to capture the transition from atomic to molecular gas at these metallicities. Therefore, in our re-simulations, we increase the resolution limits across the whole simulation box from the models in Paper I. We use the same Jeans refinement strategy, but increase the density threshold for sink particle formation to $n = 10^5$ cm$^{-3}$ and decrease the sink accretion radius and softening length accordingly. We also adopt a much smaller base gas mass of 1 M$_\odot$. The new parameters can be seen in Table 2. The resulting cell size as a function of density is shown in Figure 2. We see that at the density threshold for sink particle formation, the cell size is approximately 0.01 pc. This improved resolution and range of resolvable densities are sufficient to capture the atomic-to-molecular transition at the metallicities studied in this paper (Glover & Clark 2012).

The new models were run for 15 Myr using the snapshots at 1 Gyr from the models in Paper I as initial conditions. We choose to start from 1 Gyr as by this time the simulations are in a steady state with no large scale variations in star formation or chemical evolution for a significant number of Myr. We choose to run for an additional 15 Myr so that the models form a complete new set of sink particles, as over this timescale all the sinks formed at the previous resolution will have most likely been turned into star particles. This also ensures that the gas structures we analyse have been resolved at high resolution for their entire life cycle and that the chemistry has had time to react and settle with typical molecular cloud lifetimes being of the order a few tens of Myr (Inutsuka et al. 2015).

### 2.1.3 Star formation

Our treatment of star formation is based on an accreting sink particle model (Tress et al. 2020). Sink particles are used to replace gas cells denser than a threshold $\rho_c$ of $2.4 \times 10^{-19}$ g cm$^{-3}$ (number densities of $10^5$ cm$^{-3}$) that also satisfy the following criteria:

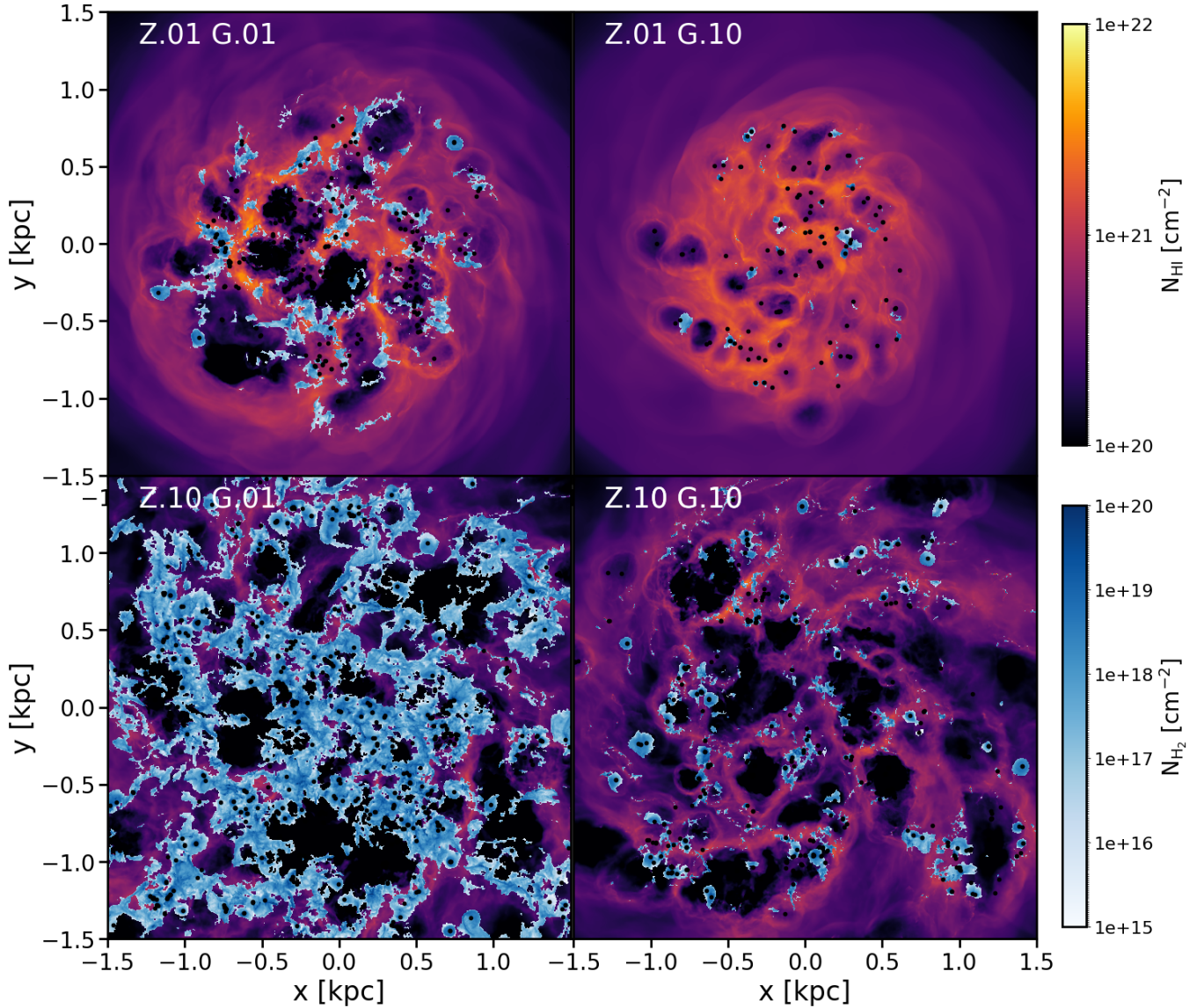

**Figure 1.** Molecular surface density ($N_{H_2}$) shown on top of HI surface density ($N_{HI}$). These are shown at 1.015 Gyr time when we begin our analysis.

(i) The gas flow must be converging with both the velocity divergence and the acceleration divergence being negative.

(ii) The region must be centred on a local potential minimum.

(iii) The region must not fall within the accretion radius of an existing sink particle.

(iv) The region should not move within the accretion radius of any other sink particle in a time less than the local free-fall time.

(v) The region within the accretion radius must be gravitationally bound.

In our new simulations, we adopt an accretion radius of $r_{acc} = 0.01$ pc for the sink particles. At this scale therefore our sinks correspond to the gas going into star formation within an individual star-forming core. However, as we do not have sufficient resolution to capture fragmentation on the scale of these individual cores, our sink particles represent stellar systems rather than individual stars (i.e. some fraction of them will be binaries, triples, higher-order multiples, or unbound associations of stars). To attribute an appropriate stellar population to each sink, we stochastically sample from a prescibred initial mass function (IMF) using the method described in Sormani et al. (2017). In Paper I the maximum mass of each sink was fixed to 200 $M_\odot$, but we here increase it to 1000 $M_\odot$ so that more massive stars are included in the stochastic sampling of the stellar IMF if the conditions require it. We also increase the star formation efficiency ($\epsilon_{SF}$) of each sink from 0.10 in Paper I to 0.32 here, i.e. 32% of each sink's mass is converted to stars, with the rest remaining as (unresolved) gas. This increase in $\epsilon_{SF}$ accounts for the greater resolution of our new simulations, which comes closer to capturing the actual star-forming scale, but also reflects the fact that even at this higher resolution, not all of the gas entering the sink is likely to end up in a star. The precise value we choose for $\epsilon_{\rm SF}$ in our new simulations is a conservative estimate, based on Matzner & McKee (2000).

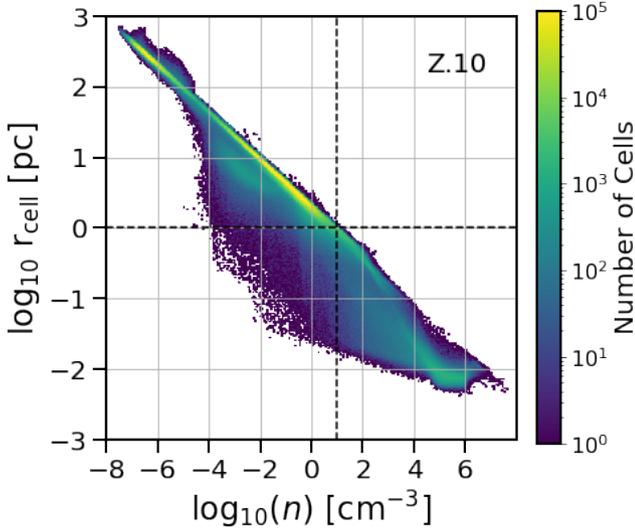

**Figure 2.** Effective cell radius, $r_{\rm cell}$, as a function of number density. The effective radius is defined as the radius of a sphere that has the same volume as the cell. The dashed black lines mark a cell size of 1 pc and a number density of n = $10 \, {\rm cm}^{-3}$. This is for model Z.10 G.10 but the results for the other models are very similar.

Once formed, a sink is allowed to accrete gas mass from surrounding gas cells that lie within $r_{\rm acc}$, have densities above $\rho_c$ and that are gravitationally bound to the sink. Cells satisfying these criteria lose sufficient gas to reduce their density to $\rho_c$, although for stability, we limit the mass accreted from a cell to 90% of its initial mass.

*2.1.4 Feedback*

In the simulations presented in this work we account for two forms of stellar feedback: supernova (SN) explosions and photoionisation of the gas by UV photons emitted by massive stars.

For every star with a mass of $\geq 8 M_\odot$ we generate an SN at the end of its lifetime. The lifetime is worked out from the stars mass in Table 25.6 of Maeder (2009). We select the location of the SN randomly within a sphere of 5 pc centered on the sink particle ($r_{\rm SNe}$). The SN injects energy or momentum into the surrounding ISM (Bubel 2015; Smith et al. 2020; Tress et al. 2020). A sphere of 32 cells around the supernova remnant defines an injection radius ($R_{\rm inj}$), which on average is ~ 28 pc. This is then compared to the SN remnant radius at the end of its Sedov-Taylor phase ($R_{\rm sr}$). If $R_{\rm sr} > R_{\rm inj}$, AREPO injects $10^{51}$ erg of thermal energy into the gas and sets the ionisation fraction to 1.0 in all of the cells within $R_{\rm inj}$. If $R_{\rm inj} > R_{\rm sr}$, the Sedov-Taylor phase is unresolved, then momentum is injected instead, the gas temperature is set to $10^4$ K and the gas fully ionised.

Along with energy or momentum, each SN also returns a fraction of the sink mass to the ISM. This accounts for the fact that the majority of the gas accreted by the sink does not form stars (since $\epsilon_{\rm SF} < 1$) and in reality would be returned to the diffuse ISM when the star-forming cloud is dispersed. To compute the mass returned by each SNe, we first determine the number of SNe associated with the sink, $N_{\rm SN}$. The ejected mass then follows as

$$M_{\rm ej} = \frac{M_{\rm sink} - M_{\rm stars}}{N_{\rm SN}}, \quad (3)$$

where $M_{\rm sink}$ and $M_{\rm stars}$ are the mass of the sink and the mass of stars formed by the sink, respectively. The ejected mass is distributed uniformly within the injection radius. A more detailed explanation can be found in Tress et al. (2020).

After the final SNe has been generated, the mass remaining in the sink particle will only be the mass of stellar objects. The sink particle is then converted into a collisionless *N*-body star particle. Sink particles that do not contain any massive stars and that hence generate no SNe are converted to star particles after 10 Myr. The mass of the star particle is based on $\epsilon_{\rm SF}$ with the remaining gas returned to the surrounding gas cells.

We also include photoionisation from massive stars. Photoionisation plays a crucial role in the evolution of the ISM because it injects energy and momentum into the surrounding medium at early times, which alters the environment in which the SNe later explode (Klessen & Glover 2016). Massive stars emit UV radiation that heats up and ionises the surrounding ISM which leads to an expansion of the gas in regions close to the star. As a consequence, when a massive star explodes as a supernova, it does so in a more diffuse region, rather than a dense molecular cloud. As a consequence, the energy and momentum generated by the explosion are less efficiently transferred to the surrounding dense gas. This leads to less disruption of the cloud and likely a reduction in the formation of dense regions (Vázquez-Semadeni et al. 2010; Walch et al. 2012; Dale et al. 2014; Sales et al. 2014; Iffrig & Hennebelle 2015). It also acts as a pressure support against the collapse of the gas surrounding the star and can prevent clustering of SNe (Smith et al. 2021). Therefore if we want to properly model the ISM and feedback from SNe on small scales then it is important that we include feedback from photoionisation.

The photoionisation model adopted in this work is a Strömgren volume approximation similar to the schemes used in Hopkins et al. (2012) and Hu et al. (2017). When a sink particle contains one or more stars of mass $M > 8$ M$_\odot$, we treat it as a source of ionising radiation. In this work we use a simplified routine for computational efficiency that is an approximation and treats the photoionisation as if produced by stars of a single mass. We find that for the scales we are studying in these models this is a good approximation, but appreciate that this is not a realistic case. We assume that each massive star emits ionising photons at a rate $10^{49}$ s$^{-1}$ and compute the total ionisation rate for the sink, $S_*$, by summing up the total number of massive stars that it contains that have not yet exploded as supernovae. Given $S_*$, we then determine the set of cells that should be ionised. To determine whether a cell should be ionised, the code walks through the cells around the sink in order of increasing radial distance (up to a maximum radial distance of 50 pc). For each cell, we determine whether the ionising flux from the sink is able to maintain it in an ionised state by computing the rate at which recombinations would occur within the cell if it were fully ionised. This is given by

$$R_c = \beta n_e^2 V_{\rm cell}, \quad (4)$$

where $\beta = 2.56 \times 10^{-13} \, \mathrm{cm^3 \, s^{-1}}$ is the case B recombination coefficient at $10^4$ K, $n_\mathrm{e}$ is the number density of electrons (assuming fully-ionised gas) and $V_\mathrm{cell}$ is the volume of the cell. This is then compared with the ionisation rate of the sink. If $S_* > R_\mathrm{c}$, the cell is flagged as ionised, the ionising flux is updated to the new value $S_{*,\mathrm{new}} = S_* - R_\mathrm{c}$, and we move on to the next cell. This process terminates once $S_* < R_\mathrm{c}$.

If a cell is flagged as ionised, its hydrogen ionisation fraction is immediately set to 1.0 and its temperature is set to $10^4$ K (if it were initially below this; for cells with initial temperature $T > 10^4$ K, we retain the original value). We prevent flagged cells from cooling below $10^4$ K and also skip the chemical update for these cells, which has the effect of disabling $H_2$ formation within them. The implementation can also cause ionised regions to overlap if sink particles are in close proximity to each other. To deal with this, once a cell is flagged as ionised it is skipped by the routine if it tries to ionise it a second time.

We note that our photoionisation routine has some limitations. Our assumption that every massive star produces ionising photons at the same rate yields roughly the right ionising flux for young stellar clusters (age $< 4$ Myr; see e.g. Leitherer et al. 1999) with a fully-sampled IMF, but substantially over-predicts the flux coming from older clusters or ones that are not massive enough to completely sample the IMF ($M < 10^4 \, \mathrm{M_\odot}$) (Cerviño et al. 2003). In our current simulations, we adopt this approach on pragmatic grounds: our main aim in including photoionisation is to ensure that any SN that explode do so in the correct environment. We therefore prefer to err on the side of making photoionsiation too effective in order to more efficiently clear the surrounding gas before the first SN.

## 3 HOW WELL DOES CO TRACE MOLECULAR GAS IN DWARF GALAXIES?

Before we investigate the star formation within the simulations we first consider how our models would appear when observed in CO. In the following analysis we restrict ourselves to a 3 kpc box containing the inner galaxy as outside of this limit we find little to no CO, especially in the Z.01 models and a rapidly decreasing molecular gas distribution.

### 3.1 CO distribution

Figure 3 shows the molecular gas distribution of the galaxies in blue. In all models substantial reservoirs of $H_2$ exist within the ISM. On top of this we plot the true density of CO Gaussian-smoothed with a kernel of 4.5 pc, reminiscent of the ALMA beam of $14.7''$ based on Band 3 observations of the SMC using a 7m baseline for $^{12}$CO J=1-0 line transition maps (Sano et al. 2019) taken from the ALMA Science Archive website (Stoehr et al. 2020) [1] looking at the Small Magellanic Cloud. Clearly, there are many $H_2$ clouds that are not traced by any CO. Figure 4 shows the same map but includes the unsmoothed CO surface density.

To quantify the mismatch between the $H_2$ and the CO, we calculate the area filling fraction for $N_\mathrm{CO}$ ($AFF_\mathrm{CO}$) and also for $N_{H_2}$ ($AFF_{H_2}$). The area filling fraction is defined here as the fractional area covered by lines of sight with $N_{H_2} \geq 10^{15} \, \mathrm{cm^{-2}}$ or $N_\mathrm{CO} \geq 10^{10} \, \mathrm{cm^{-2}}$ (in the unsmoothed map), that is how much of the projection area is covered by the gas divided by the area of the projection box ($\mathrm{Area_{proj}}$) in Figure 3, a 3 by 3 square kpc region. We choose these limits to see the distribution of $H_2$ in more diffuse gas and if any structures form and the CO that traces the densest part of the molecular gas, regions that may be seen through observations. In the CO we go to lower densities than are likely to be observed due to the sparse distribution in the discs:

$$AFF_{H_2} = \frac{\mathrm{Area}_{N_{H_2}} \geq 10^{15} \mathrm{cm}^{-2}}{\mathrm{Area_{proj}}} \, , \quad (5)$$

$$AFF_\mathrm{CO} = \frac{\mathrm{Area}_{N_\mathrm{CO}} \geq 10^{10} \mathrm{cm}^{-2}}{\mathrm{Area_{proj}}} \, . \quad (6)$$

[1] https://almascience.eso.org/aq/

| Species | Z.10 G.10 | Z.10 G.01 | Z.01 G.01 | Z.01 G.10 |
|---|---|---|---|---|
| $H_2$ | 0.109 | 0.501 | 0.085 | 0.011 |
| CO | 0.007 | 0.108 | 0.003 | 0.00022 |
| Ratio | 0.067 | 0.215 | 0.031 | 0.019 |

**Table 3.** Area filling fraction of $H_2$ and CO in the different simulations, defined as the fraction of the area with a surface density greater than $10^{15} \, \mathrm{cm^{-2}}$ (for $H_2$) or $10^{10} \, \mathrm{cm^{-2}}$ (for CO). In the final row of the table, we show the ratio between the two filling fractions, $CO/H_2$. This is extremely sensitive to the local shielding environment and varies significantly between the models.

Table 3 shows the area filling fractions for the different models. The filling fraction for $H_2$ varies from 1% to 50% of the studied disc area for the models, but for CO it is far smaller, being only 0.02% to 10%. The relative filling fraction of H2 and CO varies greatly between the models demonstrating how sensitive the CO is to the local shielding and ISRF. As the ratio of the area filling fraction of CO to $H_2$ is below 0.5 in all cases it is clear that the majority of area is likely to be CO dark. To see what the relationship is in a broader context we also plot $N_\mathrm{CO}$ against $N_{H_2}$ and the normalised fractional abundances for $H_2$ (red line) and CO (blue line) in Figure 5. We define the normalised fractional abundance as $f_i = 1$ for the peak abundance of species $i$, We see little variability in the $N_\mathrm{CO}$ / $N_{H_2}$ ratio, especially in the densest gas, with all models having a slope of $\sim 1.1$. This linear relation is different to that seen by Hu et al. (2022) who report clear differences between three different regimes. This is likely to come from the fact that we sample the entire galactic disc, which covers many different environments, whereas Hu et al. (2022) only focus only a single cloud. In all models we see $H_2$ forming at lower densities than CO.

### 3.2 Identifying regions

The next question, is whether, for the regions where both CO and $H_2$ are present, the CO emission is observable. Radiative transfer post-processing of the entire disk at the resolution necessary to resolve individual CO-bright cores is too computationally expensive to be practical. Furthermore, given the results above, it is also unnecessary, since CO is absent from the majority of the galactic disc. Instead we look at a set of 100 pc$^3$ regions in each model. This allows us to capture multiple scales of structures, from filaments to giant molecular clouds, and to fully sample the diffuse, CO-dark gas connected to those structures.

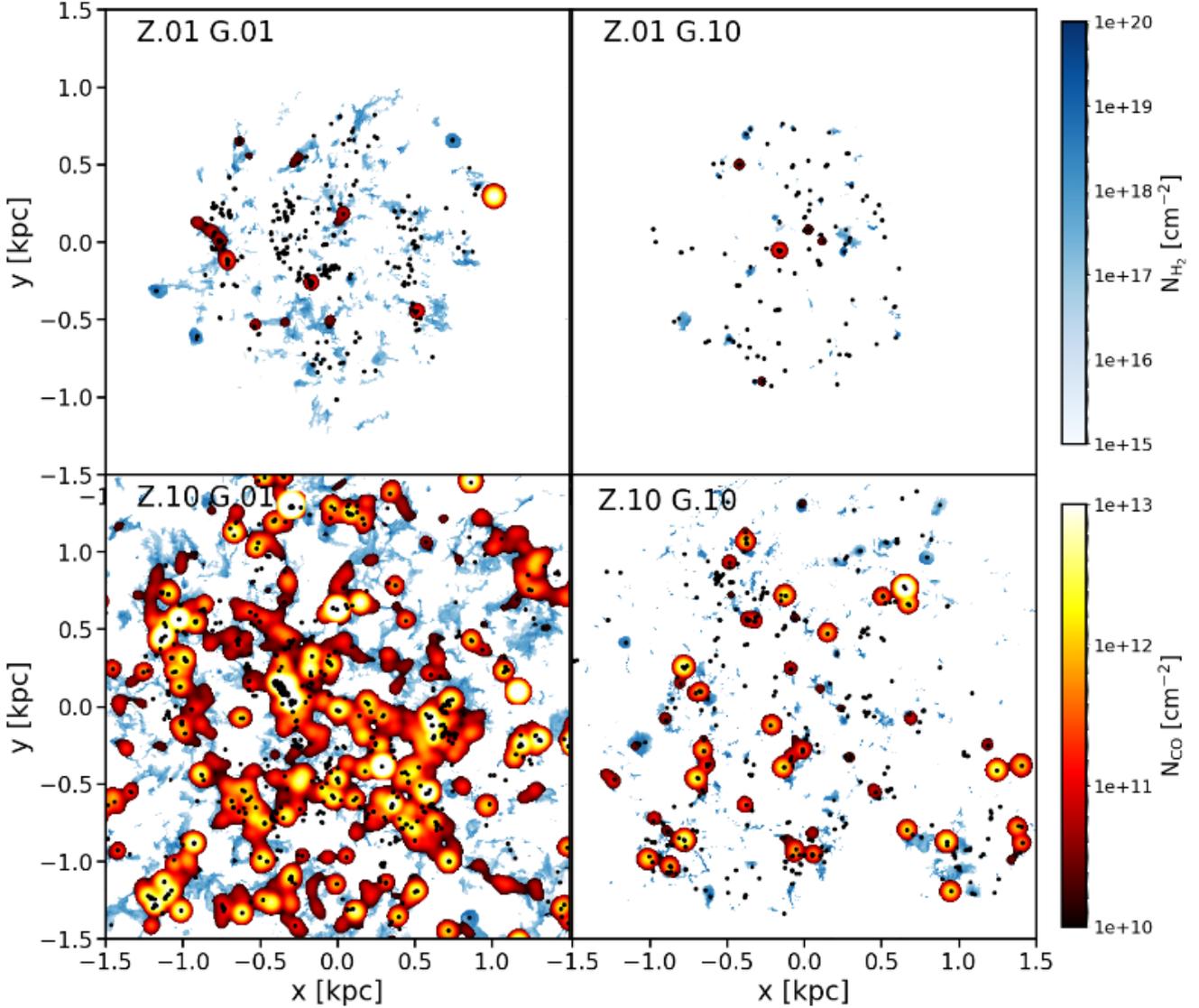

**Figure 3.** Gaussian-smoothed CO surface density ($N_{CO}$) to 4.5pc shown on top of an unsmoothed map of $H_2$ surface density ($N_{H_2}$). The black dots represent sink particles. We can see that in all models there are regions of $H_2$ that are not traced well or at all by CO. When we vary the interstellar radiation field (ISRF) independently of the metallicity there are large changes in the CO distribution, most notable between the two models on the left, Z.10 G.10 and Z.10 G.01, where the bottom plot has an ISRF an order of magnitude lower.

In order to identify regions to study we take the CO surface density projection after 15 Myr. We smooth the projection with a 2D Gaussian kernel with a standard deviation ($\sigma$) set to 10 pc, which is on the scale of beams used to observe molecular clouds in the SMC (Chevance et al. 2020a). By using astrodendro[2], we define regions of CO that have surface densities exceeding $10^{10}$ cm$^{-2}$, with each level in the dendrogram representing an order of magnitude increase in surface density. We also ensure that each level includes at least 10 pixels, where each pixel in the projection is (5 pc)$^2$. This allows us to identify various different structures within the gas. We show the regions selected from each model in Appendix A.

Once we have defined the CO surface density distribution using the dendrograms we choose each region based on the peak Gaussian smoothed density within the CO structure, selecting each region so that the peak smoothed surface density sits at the centre of a 100 pc box. This allows us to find a variety of different structures and CO distributions from each model with peak surface densities ranging from $10^{10}$ cm$^{-2}$ to $10^{15}$ cm$^{-2}$. We do this to be reflective of the range of environments within the models. It can be seen in Figures 3 and

[2] http://www.dendrograms.org

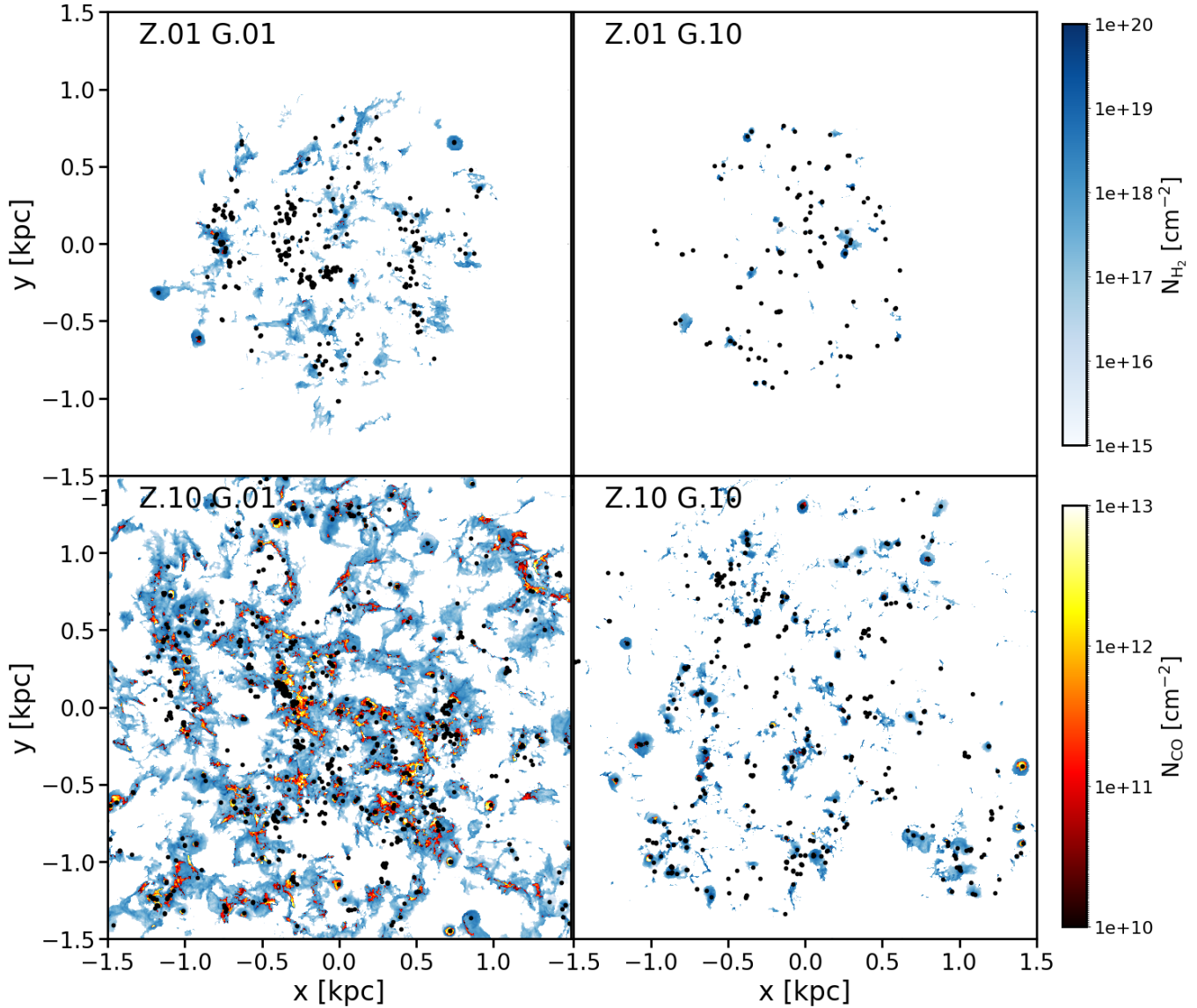

**Figure 4.** CO surface density ($N_{CO}$) shown on top of the H$_2$ surface density ($N_{H_2}$). The black dots represent sink particles. We can see that in all models there is very little CO and need the beam to smooth the CO so we can visualise it.

4 that CO exists in many different parts of the galaxy and does not always seem to trace the molecular gas well.

### 3.3 Radiative transfer post-processing

To calculate the dark gas fraction ($f_{DG}$) of the regions we perform radiative transfer using POLARIS[3] (Polarized Radiation Simulator, see Reissl et al. 2016 and Brauer et al. 2017 for further details). We use the converter from Izquierdo et al. (2021) to setup the POLARIS simulations using the same mesh as the AREPO snapshot to avoid degrading our resolution. We perform this step for each of the regions chosen above, though we note that this does *not* result in a global, disc averaged,

($f_{DG}$) as we are excluding large volumes of the diffuse molecular gas. Looking at Figure 3 we can see there are multiple H$_2$-rich regions that are not associated with CO, even when looking at the smoothed CO distribution.

POLARIS solves the radiative transfer equation for line emission and has to estimate the level populations. For this work we use the Large Velocity Gradient (LVG) approximation. In this approximation POLARIS calculates the level population for each cell in the grid using only local properties, under the assumption that line photons emitted by more distant gas are Doppler shifted out of the line and hence do not interact. To produce the synthetic emission maps we use the $^{12}$CO J=1-0 ground state rotational transition. When calculating collisional excitation and de-excitation rates, we assume that collisions with H$_2$ dominate, since most of the CO is found in H$_2$-rich gas. We use collisional rate coefficients, radiative transition probabilities and energy levels from the

[3] http://www1.astrophysik.uni-kiel.de/~polaris

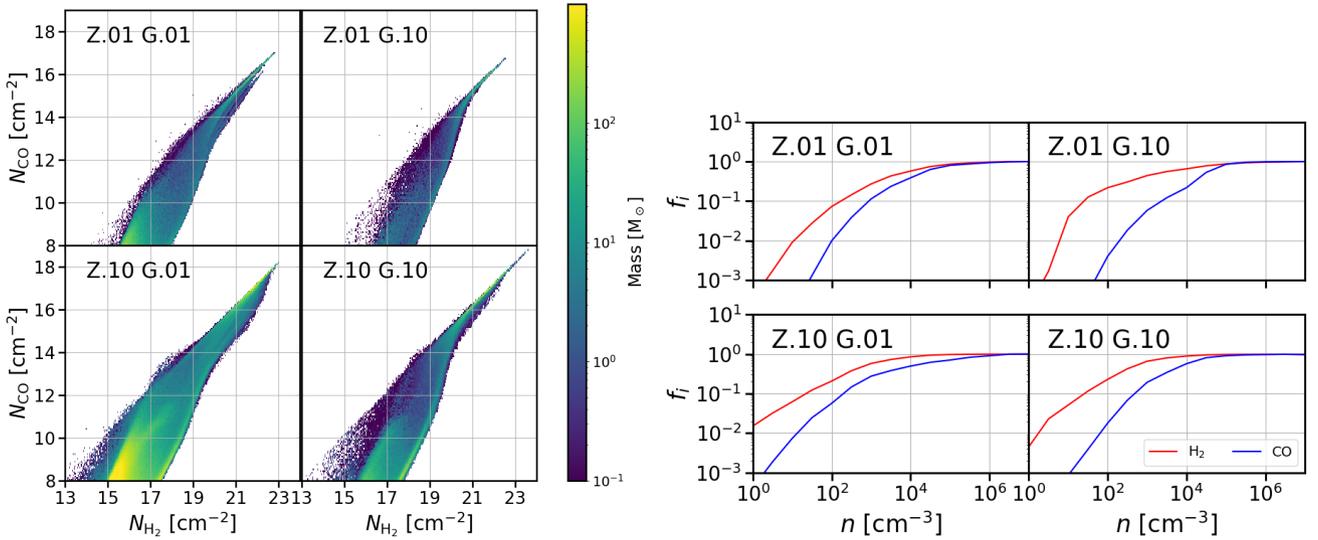

**Figure 5.** Left: The mass-weighted number density of H$_2$ ($N_{\rm H_2}$) and CO ($N_{\rm CO}$) for each model. The black line shows the fitted slope to the data which is $\sim 1.1$ for each model. Right: The normalised fractional abundance of H$_2$ and CO for each model in relation to total gas number density.

Leiden Atomic and Molecular Database (LAMDA) (Schöier et al. 2005). Using these, we run POLARIS on each of the 48 selected regions as defined in Section 3.2 to calculate a position-position-velocity (PPV) cube of line emission for that region. Each PPV cube has 67 velocity channels with a channel width of $\Delta v = 0.6\,\rm km\,s^{-1}$. This is motivated by the work of Glover & Smith (2016) who show that the average sound speed of CO-dark molecular gas in simulated Milky Way-like galaxies is $c_s = 0.64\,\rm km\,s^{-1}$; we assume that the sound speed in the lower metallicity systems studied here is unlikely to be smaller than this value. We also use a Line of Sight (LoS) velocity range of $v_{\rm min} = -20$ to $v_{\rm max} = 20\,\rm km\,s^{-1}$, as outside of this range there is very little CO and extending the considered range beyond this has a minimal effect on the results. After converting the PPV cubes to temperature and velocity units, we then integrate with respect to velocity to derive the integrated intensity of $^{12}$CO J=1-0, $W_{\rm CO}$.

To simplify comparisons to observations, we set the distance to the source from the observer to be 62.1 kpc, the distance to the Small Magellanic Cloud (SMC; see Graczyk et al. 2014). We compare to the SMC due to its similar mass, $4.2 \times 10^8\,\rm M_\odot$ (Harris & Zaritsky 2004), and metallicty, $Z = 0.1\,\rm Z_\odot$ (Harris & Zaritsky 2004), to our fiducial model, where our total gas mass is $\sim 8 \times 10^7\,\rm M_\odot$. Since the SMC is our closest metal poor companion this represents a best case scenario for the observation of CO in such environments. The full selection of regions upon which radiative transfer was performed and their calculated CO emission is shown in the Appendix in Figure A1.

### 3.4 CO-dark gas

To compute the CO-dark fraction of each region in the subsample we define:

$$f_{\rm DG} = \frac{M_{\rm H_2}^{W_{\rm CO}}}{M_{\rm H_2}^{\rm Tot}}, \qquad (7)$$

where $M_{\rm H_2}^{W_{\rm CO}}$ is the mass of H$_2$ along lines of sight that have integrated intensities $W_{\rm CO} < 0.1\,\rm K\,km\,s^{-1}$ and $M_{\rm H_2}^{\rm Tot}$ is the total mass of H$_2$ within the 100 pc$^3$ region, which is easily measured using the underlying simulation. Note that in our calculation of $M_{\rm H_2}^{W_{\rm CO}}$, we ignore any contribution by gas outside of the 100 pc$^3$ region, to ensure we can make a meaningful comparison with $M_{\rm H_2}^{\rm Tot}$.

In Figure 6 we plot the cumulative H$_2$ mass as a function of $W_{\rm CO}$ for each model at two different beam sizes. We choose two different beam sizes based upon ALMA observations of the Small Magellanic Cloud (SMC), the 14.7″ as discussed in Section 3.1 as well as a smaller 1.5″ beam based on a 12m baseline for $^{12}$CO J=1-0 line transition map (Sano et al. 2019). We use ALMA beam sizes as we are synthesising SMC-like CO observations where the peak intensities are almost point-like sources. We note that our synthetic images are not the same as what one would actually observe using an interferometer, as we have not accounted for interferometric filtering of large-scale emission. However, in practice, this simplification is likely to have little impact on our results, as there is little extended bright CO emission in any of these systems. The black line in each panel in Figure 6 represents the mass-weighted average relationship between cumulative H$_2$ mass and $W_{\rm CO}$ for each simulation and beam size with the blue and red lines representing each individual region. The values for each region are shown in full in the Appendix A, but we summarise the averages here in Table 4.

We see that when the CO emission is smoothed with a large beam, very few of our analyzed regions have detectable CO emission. In simulation Z.10 G.10, all of the selected regions have a CO integrated intensity $\ll 0.1\,\rm K\,km\,s^{-1}$, and the same is true for almost all of the regions in simulation Z.01 G.01. In simulation Z.01 G.10, a few regions have integrated intensities slightly above our detection threshold, although the average over all regions lies below the threshold. In all three of these simulations, the dark gas fraction is therefore $\sim 100\%$, i.e. CO observations will miss essentially all of the H$_2$ in these systems. The only case in which significant amounts of detectable CO emission is present is simulation Z.10 G.01, and even in this case many of the selected regions are CO-dark and the dark gas fraction is large ($f_{DG} = 0.55$).

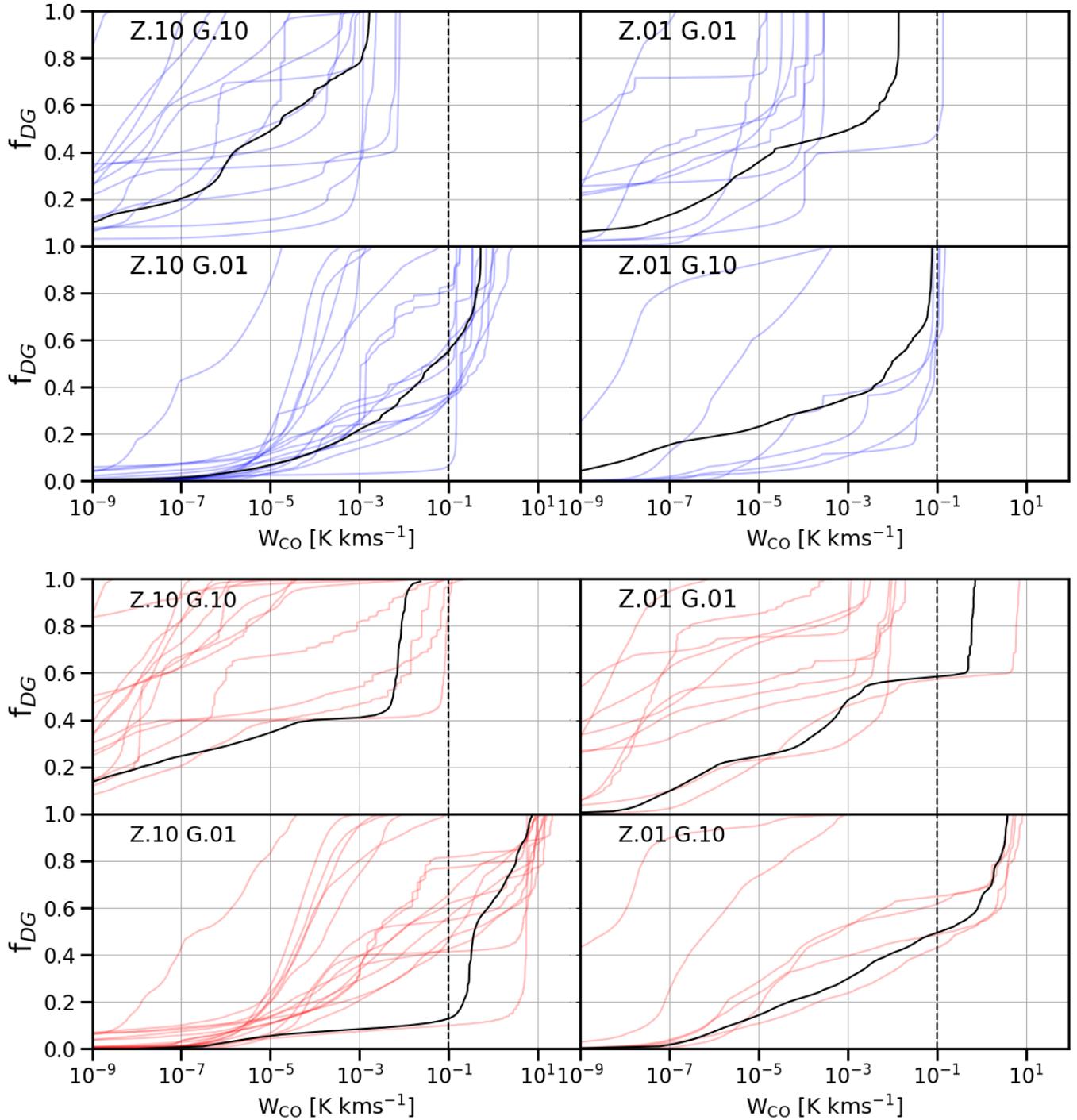

**Figure 6.** Cumulative fraction of the total H$_2$ mass located in regions with W$_{CO}$ less than or equal to the listed value, plotted as a function of W$_{CO}$ for all models. The top 4 plots with blue lines are for when observed with a 14.7″ beam. The bottom 4 plots with red lines are for a 1.5″ beam. The solid black line shows the mass-weighted average of the regions for each model and the dashed line shows a value of W$_{CO}$ = 0.1 K km s$^{-1}$. We consider molecular gas located in regions with CO integrated intensities smaller than this value to be CO-dark.

Smoothing the CO map with a smaller beam, which better matches the actual size of the CO-rich cores, makes it easier to find regions with detectable CO emission, particularly in runs Z.10 G.01 and Z.01 G.10. However, even in this case, many regions remain undetectable and the dark gas fraction remains high, ranging from 60–100%. In all cases, even in the rare cases where CO is present within the molecular gas, it is unobservable for the majority of regions. When we look at the mass-weighted average for the regions we find that model Z.10 G.01 has much of its molecular mass in regions that are CO-bright according to the definition used here, although still considerably fainter than would be typical for GMCs in the Milky Way. We note, however, that this is an average over our selected regions and not the galaxy as a whole, and hence may be biased towards regions with higher $W_{\rm CO}$.

| Model | 14.7″ | 1.5″ |
|---|---|---|
| Average Z.10 G.10 | 1.00 | 1.00 |
| Average Z.10 G.01 | 0.49 | 0.13 |
| Average Z.01 G.01 | 1.00 | 0.59 |
| Average Z.01 G.10 | 1.00 | 0.50 |

**Table 4.** Dark gas fractions ($f_{\rm DG}$) for the two beam sizes across all regions within the sub-sample of representative regions where CO is present. We can see that $f_{\rm DG}$ is high for all models.

We also look at the $X_{\rm CO}$ conversion factor for each region; see Appendix A1 for details of how we calculated this. We find that there is large variability in the value of $X_{\rm CO}$ in regions where we have CO-bright gas. This is in line with both recent observational (Ramambason et al. 2023) and numerical (Hu et al. 2022) results.

## 4 THE STAR-FORMING ENVIRONMENT

The results presented in the previous section demonstrate that the majority of the molecular gas in our simulated low metallicity dwarf galaxies is CO-dark. However, we also know that these galaxies are star-forming, despite their lack of observable CO, as shown in Paper I. It is therefore clear that these galaxies will deviate from the $L_{\rm CO}$–star formation rate correlation observed in metal-rich spirals, in line with what is observed in real dwarf galaxies (Schruba et al. 2012). However, there remains the question of whether stars in these systems form in clouds that are dominated by $H_2$ (i.e. in CO-dark molecular gas), or whether their birth environment is dominated by cold atomic gas. In this section, we use the results of our simulations to attempt to distinguish between these two scenarios.

### 4.1 Sink age versus density

At the resolution achieved in our resimulations ($\sim 0.01$ pc at the density threshold for sink formation), it is reasonable to take each sink particle to represent an individual protostellar system (i.e. a single protostar or a small-$N$ multiple). We can therefore use the state of the gas surrounding each sink as a measure of its local environment.

To put this on quantitative terms, we first define a cubical region of 3 pc diameter around each sink for analysis (see Appendix B for a similar analysis for different region sizes). The size of this region is large enough to encompass not only the gas gravitationally bound to the sink, but also a surrounding envelope of more diffuse material. If we take a smaller region we risk losing information on the diffuse envelope, whilst adopting a larger region risks biasing our results for sinks that sit on boundaries between cold, dense gas and hot diffuse gas. By taking a size of 3 pc we have limited that effect as much as possible, although we note that there are still a few sinks on these boundaries. We also note that our sink accretion radius is 0.01 pc and that by going much further out we are looking at regions that may well not be involved in the star formation process in the models.

We have measured the mass and volume-weighted $H_2$ number density and $H_2$ fraction within each of the $(3\,{\rm pc})^3$ regions surrounding the sinks at the end of the resimulations, and show the resulting values as a function of sink age in Figures 7 and 8. Note that our high resolution simulations span only a short period in the lifetime of the dwarf galaxies and therefore the new sinks formed in them are all young, with ages of up to $\sim 10$ Myr. We discuss the choice between mass-weighted and volume-weighted averaging in Appendix C and present both here for completeness.

We see from Figures 7 and 8 that the molecular gas density surrounding the youngest sinks (age $< 2$ Myr) is very high. The mass-weighted $H_2$ number density is $\sim 10^5\,{\rm cm}^{-3}$ around all of the young sinks. The volume-weighted average is much lower, $\sim 100\,{\rm cm}^{-3}$, indicating that the dense, $H_2$-rich gas only occupies a small fraction of the analysed volume. The mean $H_2$ abundance within each $(3\,{\rm pc})^3$ volume is also large, $x_{H_2} \simeq 0.5$, showing us that this gas is very close to being fully molecular. On these small scales, there is hence little evidence for stars forming in regions dominated by atomic gas. As shown by Glover & Clark (2012), it is the density of gas in the environment that drives the star formation, not specifically molecular gas. If stars form they must do so in dense gas, and the formation of this gas also allows $H_2$ to form in substantial quantities. At the high densities found around the sink particles, this occurs quickly even in our lowest metallicity simulations.

Interestingly, sinks older than $\sim 2$ Myr start to appear in diffuse gas ($n_{H_2} \leq 100\,{\rm cm}^{-3}$, for both mass and volume-weighting). In other words, the older the sink, the greater the likelihood that it is found in diffuse gas with a low $H_2$ fraction. This seems to confirm the idea of a disconnect between young stars and dense gas (Kruijssen et al. 2018; Schinnerer et al. 2019; Chevance et al. 2020a) that occurs soon after the onset of star formation. But how do the sinks arrive in the diffuse gas and what mechanism has caused the formation of these diffuse regions? There are several possibilities. The first is feedback. Supernovae from sinks begin to detonate at times $\sim 2.7$ Myr after formation for the most massive stars. This coincides well with the age at which we start to see decoupling and may indicate that the molecular gas is easily destroyed by feedback.

Table 5 shows the fraction of sinks that exist at different ages and densities. We see that in 3 of the 4 models most of the sinks are found in dense gas (i.e. regions with volume-averaged $H_2$ number densities greater than $100\,{\rm cm}^{-3}$. The exception is run Z.10 G.01, in which a majority of sinks are found in low density regions. Interestingly, this is also the simulation that forms the most stars (see Whitworth et al. 2022). The relatively high number of sinks found in diffuse gas in this simulation may therefore be due to stellar feedback clearing the gas away from the sinks more effectively than in the lower star formation rate simulations.

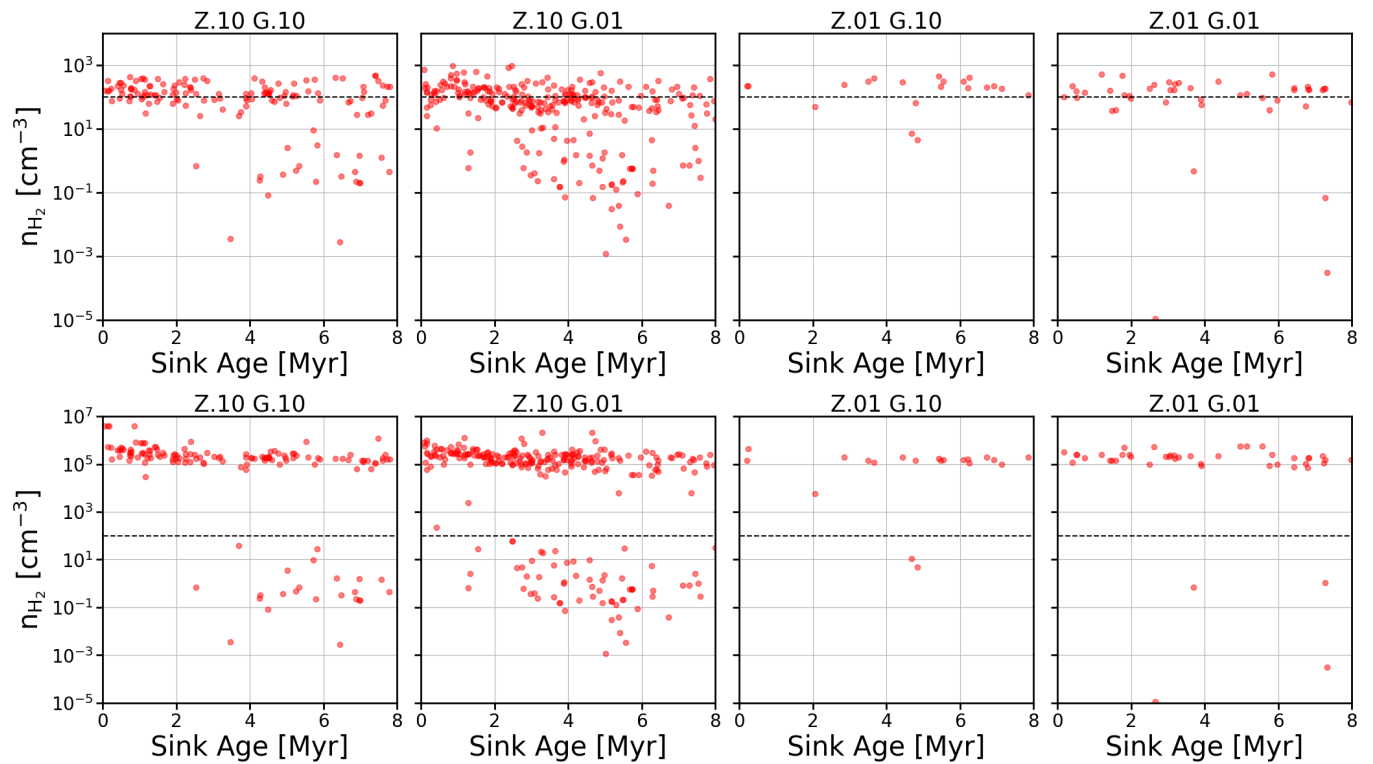

**Figure 7.** Top: The volume-weighted number density of $H_2$ in a $(3\,\mathrm{pc})^3$ cube centred on a sink plotted in relation to the age of the sink. Bottom: The mass-weighted number density of $H_2$ in a $(3\,\mathrm{pc})^3$ cube centred on a sink plotted in relation to the age of the sink. The dashed line represents the turn over to dense, cold gas at a number density of $100\,\mathrm{cm}^{-3}$.

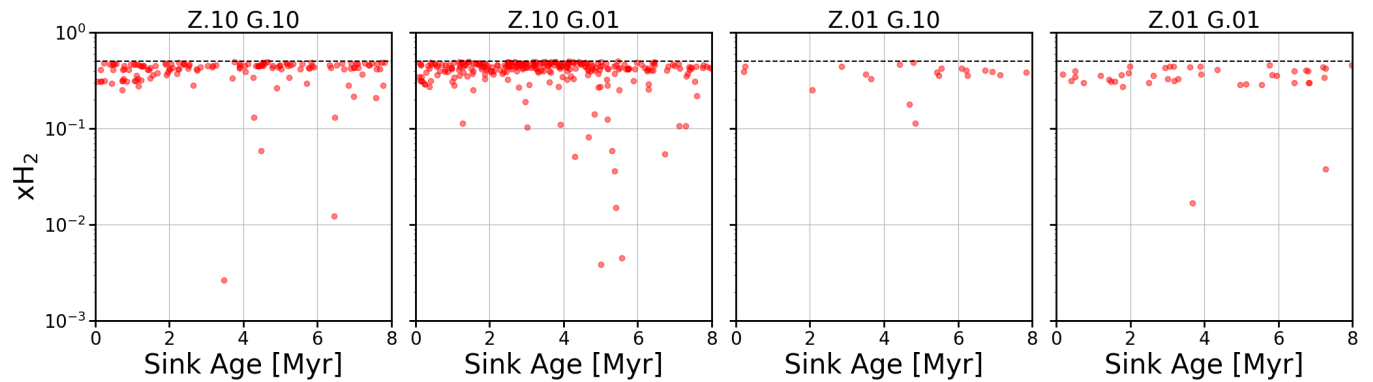

**Figure 8.** The mean fractional abundance of $H_2$ in a $(3\,\mathrm{pc})^3$ cube centred on a sink plotted in relation to the age of the sink. The dashed line indicates the value $x_{H_2} = 0.5$ that corresponds to fully molecular gas.

If we overplot the location of the sink particles on top of the $H_2$ and CO column densities on Figure 4, we see that most of the sinks are located in dense molecular regions. To see whether the regions around the sinks remain cold and dense over time, we plot the temperature-density phase diagram of the combined models for the $(3\,\mathrm{pc})^3$ cube around the sinks at 1 Myr age intervals in Figure 9. We can see that most of the gas has densities $n \geq 100\,\mathrm{cm}^{-3}$ and temperatures $T \leq 100\,\mathrm{K}$ (black lines in the plots). The only variation we see in the dense cold gas over time is a reduction in its total mass. This is likely due to a combination of accretion onto the sink (removing the densest, coldest gas) and feedback (removing the lower density, warmer material). The horizontal line in temperature at $10^4\,\mathrm{K}$ is the supernovae and photoionisation injection temperature.) If we look at the difference in total gas mass and cold gas mass around different aged sinks in all models, Table 6, we see a large reduction in the total gas mass as sinks get older. For the cold gas we see a decrease of a factor of $\sim 2.3$, which is slightly more than the increase in the sink mass. Therefore, cold gas mass loss is most likely driven by accretion onto the sinks, whilst total mass loss is driven by feedback with some cross over. We see that almost all of the gas around the sinks is cold, and that a substantial fraction is molecular, even for older sinks. We carry out a similar analysis for larger regions, (5 pc)$^3$ and (10 pc)$^3$, in Appendix B and see similar trends in the change in mass, though the larger the cube, the smaller the ratio between molecular mass and total gas mass.

| Model | Fraction $n_{H_2} \geq 100$ cm$^{-3}$ | Fraction < 2 Myr & $n_{H_2} \geq 100$ cm$^{-3}$ | Fraction > 2 Myr & $n_{H_2} \geq 100$ cm$^{-3}$ | Fraction < 2 Myr & $n_{H_2} \leq 100$ cm$^{-3}$ | Fraction > 2 Myr & $n_{H_2} \leq 100$ cm$^{-3}$ |
|---|---|---|---|---|---|
| Z.10 G.10 | 0.58 | 0.30 | 0.28 | 0.06 | 0.37 |
| Z.10 G.01 | 0.34 | 0.19 | 0.15 | 0.09 | 0.57 |
| Z.01 G.10 | 0.82 | 0.07 | 0.75 | 0.00 | 0.18 |
| Z.01 G.01 | 0.57 | 0.25 | 0.32 | 0.15 | 0.28 |

**Table 5.** Fraction of sinks in each model that have a volume-weighted molecular gas number density greater than 100 cm$^{-3}$ (column 1), are younger than 2 Myr with a molecular gas number density greater than 100 cm$^{-3}$ (column 2), are older than 2 Myr but a molecular gas number density of greater than 100 cm$^{-3}$ (column 3), are younger than 2 Myr with a molecular gas number density greater than 100 cm$^{-3}$ (column 4), and are older than 2 Myr with a molecular gas number density less than 100 cm$^{-3}$ (column 5). We see that although less likely, some young sinks do exist in diffuse gas.

| Sink Age (Myr) | Total gas (M$_\odot$) | Total H$_2$ (M$_\odot$) | T ≤ 300 K (M$_\odot$) | Total sink (M$_\odot$) |
|---|---|---|---|---|
| 1 | 44219 | 29572 | 43178 | 6696 |
| 2 | 26860 | 18496 | 25677 | 9084 |
| 3 | 16950 | 12293 | 16145 | 13309 |
| 4 | 13521 | 9165 | 12110 | 13287 |
| 5 | 14185 | 11646 | 13260 | 20799 |
| 6 | 8306 | 6251 | 7891 | 15134 |
| 7 | 8891 | 6331 | 8325 | 13278 |
| 8 | 9052 | 7269 | 8481 | 14542 |

**Table 6.** The total gas mass, molecular gas mass, gas mass below 300 K surrounding sinks of different ages in all models compared to the total mass in the sinks at those ages. The sink mass increases by a factor of ∼ 2 from the youngest to oldest, while the total gas mass decreases by a factor ∼ 5, the H$_2$ mass by a factor of ∼ 4, and the cold gas mass by a factor of ∼ 5.

These results are consistent with a picture in which the stars in our simulations are forming in dense, cold, highly molecular cores that have small physical sizes and that are surrounded by less dense, predominantly atomic gas. For example, in the individual cloud models of Glover & Clark (2012), the dense star-forming cores are fully molecular at densities of ∼ 10$^5$ cm$^{-3}$ and above, even at metallicities as low as 0.01 Z$_\odot$, but the composition of the gas on the 10–20 pc cloud scale is a strong function of metallicity, ranging from highly molecular at solar metallicity to almost completely atomic at low metallicities. The more recent, larger-scale simulations by Hu et al. (2022) and Hu et al. (2023) also largely agree with this picture, finding high molecular fractions at high densities for all metallicities 0.1 ≤ Z$_\odot$ ≤ 3.0, but a composition at low densities that is a much stronger function of metallicity.

Archer et al. (2024) show observationally that in an irregular dwarf galaxy objects, defined as FUV bright star-forming regions, that are closer to CO-bright regions tend to be more massive. Looking at Figure 10 we note that in our disc-like dwarf galaxy this is not the case. Most of our sinks sit in or close to the dense gas. There appears to be no correlation in any model between the sink mass and the properties of the dense molecular gas, although it should be noted that our sinks are only probing a relatively limited range in system mass. Further investigation of this point is a topic for future work.

## 5 CAVEATS

Several points should be borne in mind when considering the results of this study. First, there are two important points to note regarding the analysis of our results. The first is that as the time followed in our high resolution simulation is much shorter than the H$_2$ formation time in low density gas, it is possible that the H$_2$ fraction will not have reached chemical equilibrium there, especially if the gas is well shielded or at low G$_0$. Second, our selecting of regions for further analysis and radiative transfer post-processing is informed by their CO content and hence is inevitably biased towards regions that have non-negligible CO. Care should therefore be taken when extrapolating from these results to the behaviour of the galaxy as a whole.

Second, our treatment of photoionisation, which assumes (reasons of computational efficiency) that all massive stars produce ionising photons at a constant rate of 10$^{49}$ photons per second, will tend to over-estimate the impact of photoionisation, particularly from older star-forming regions or ones deficient in the most massive stars.

Third, magnetic fields are not included in these models. The inclusion of these affects the formation time and size of the molecular clouds due to suppression in the collapse of the cloud as has been shown by Girichidis et al. (2018) and Whitworth et al. (2023). However, their impact on the sub-cloud scales that we are primarily interested in here is likely to be smaller.

Finally, for simplicity we have assumed a uniform far-UV interstellar radiation field. This assumption obviously will break down close to regions of massive star formation, and hence we will underestimate the effects of photoelectric heating in these regions. However, previous simulations of star formation in low metallicity dwarf galaxies suggests that photoionisation and supernovae, both of which are included in our model, are far more important sources of feedback than photoelectric heating (Hu et al. 2017). Our results should therefore be relatively insensitive to this simplification.

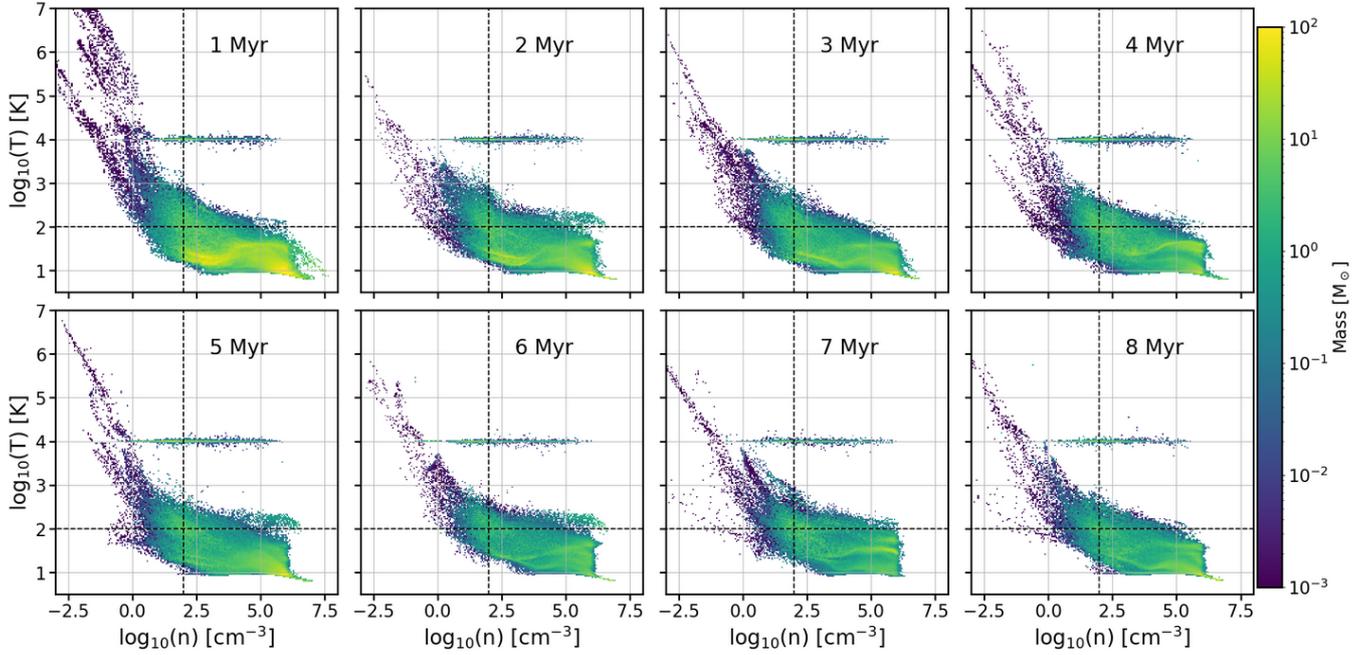

**Figure 9.** Temperature-density phase plots of the $(3\,\mathrm{pc})^3$ cube around the sinks in each model. We make a plot for sinks at different ages. The dashed lines highlight a density of $n = 100\,\mathrm{cm}^{-3}$ (vertical line) and a temperature of $T = 100\,\mathrm{K}$ (horizontal line), which roughly bounds the cold, dense, star-forming gas. The horizontal distribution of points at $T \sim 10^4\,\mathrm{K}$ corresponds to photoionized gas in the HII regions surrounding the massive stars.

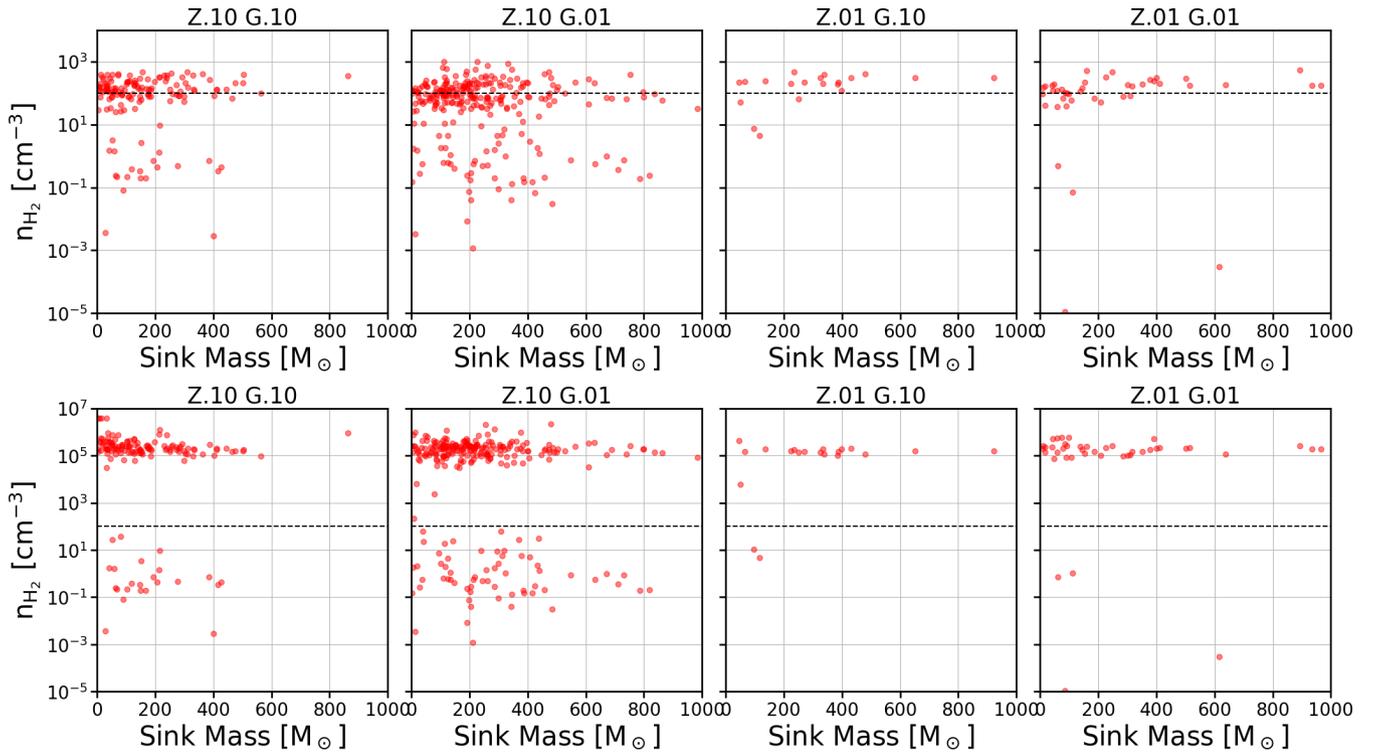

**Figure 10.** As Figure 7, but showing the volume-weighted (top) and mass-weighted (bottom) $H_2$ number densities as a function of the mass of the sink.

# 6 CONCLUSIONS

In this work we have presented four ultra high resolution hydrodynamical simulations of dwarf galaxies that resolve physical scales as small as 0.01 pc in the densest gas, and follow the evolution of the gas up to densities as high as $10^6\,\mathrm{cm}^{-3}$. These simulations include photoionisation and SNe feedback, as well as a time-dependent non-equilibrium chemical network. We studied the the dark gas fraction in each model and how the star-forming sink particles relate to the gas density over time.

From the study of the gas density around the sink particles we can see that over time a separation develops between the location of sinks and dense gas in the ISM. Some sinks stay in dense gas, and some exist in much more diffuse gas. However, the youngest sinks do always exist within cold and dense gas, but due to the high dark gas fraction they will be hard to observe.

Below we list our main conclusions.

(i) Sink particles (representing star-forming regions) initially form in clumps of cold, dense, molecular gas even at metallicities as low as $0.01\,\mathrm{Z}_\odot$. These clumps have small physical sizes ($< 1$ pc) and are surrounded by lower density atomic and molecular gas.

(ii) These dense clumps of gas can in some cases have appreciable CO luminosities, although there is considerable region-to-region variation. However, their small physical sizes mean that observations probing physical scales much larger than the clumps will be strongly affected by beam dilution and will recover very low CO integrated intensities.

(iii) For this reason, the fraction of CO-dark molecular gas in these galaxies is very large, particularly in the simulations with very low metallicity or a high interstellar radiation field. Systems such as these would therefore be difficult to observe using the standard observational proxies for $H_2$.

(iv) At early times ($t < 2$ Myr following their formation), sinks are found almost exclusively in dense molecular gas. At later times, a growing number decouple from the dense gas, due to the combined influence of ongoing star formation and stellar feedback.

# ACKNOWLEDGEMENTS

We thank the anonymous referee for a thorough and insightful report that helped us sharpen and clarify the arguments in this paper. We also thank the members of the AREPO ISM group and ECOGAL Large Scale Structures group for discussions and insightful comments on the coding and science goals in this paper. We thank Volker Springel for giving us access to the development version of AREPO. This project was supported by computing time from the STFC under the DiRAC thematic project ACTP247. DJW acknowledges support from the Programa de Becas posdoctorales of the Direcci ón General de Asuntos del Personal Académico of the Universidad Nacional Autónoma de México (DGAPA,UNAM,Mexico). DJW also thanks the PhysISM group and Roberto Galvin at the Instituto de Radioastronomía y Astrofísica at UNAM for discussions and comments on the work. RJS gratefully acknowledges an STFC Ernest Rutherford fellowship (grant ST/N00485X/1).

DJW, SCOG, NB, and RSK acknowledge funding from the European Research Council via the ERC Synergy Grant "ECOGAL" (project ID 855130), from Deutsche Forschungsgemeinschaft (DFG) via the Collaborative Research Center "The Milky Way System" (SFB 881 – funding ID 138713538 – subprojects A1, B1, B2 and B8), from the German Excellence Strategy via the Heidelberg Cluster of Excellence "STRUCTURES" (EXC 2181 - 390900948), and from the German Ministry for Economic Affairs and Climate Action in project "MAINN" (funding ID 50OO2206). They also acknowledge computing resources provided by *The Länd* and the DFG through grant INST 35/1134-1 FUGG and data storage at SDS@hd through grant INST 35/1314-1 FUGG. NB also acknowledges support from the ANR BRIDGES grant (ANR-23-CE31-0005). RSK also thanks the Harvard-Smithsonian Center for Astrophysics and the Radcliffe Institute for Advanced Studies for their hospitality during his sabbatical, and the 2024/25 Class of Radcliffe Fellows for highly interesting and stimulating discussions. PCC acknowledges the support of a consolodated grant (ST/K00926/1) from the UK Science and Technology Funding Council (STFC).

This work used the DiRAC COSMA Durham facility managed by the Institute for Computational Cosmology on behalf of the STFC DiRAC HPC Facility (www.dirac.ac.uk). The equipment was funded by BEIS capital funding via STFC capital grants ST/P002293/1, ST/R002371/1 and ST/S002502/1, Durham University and STFC operations grant ST/R000832/1. DiRAC is part of the National e-Infrastructure. The research conducted in this paper used SciPy (Virtanen et al. 2020), NumPy (van der Walt et al. 2011), and matplotlib, a Python library used to create publication quality plots (Hunter 2007). It also made use of astrodendro, a Python package to compute dendrograms of Astronomical data (http://www.dendrograms.org).

# DATA AVAILABILITY

The integrated emission data is uploaded to Zenodo.org via the link: https://doi.org/10.5281/zenodo.13930745. The full data cube outputs from POLARIS are availably upon request to the first author.

# REFERENCES

Archer H. N., Hunter D. A., Elmegreen B. G., Rubio M., Cigan P., Windhorst R. A., Cortés J. R., Jansen R. A., 2024, arXiv e-prints, p. arXiv:2404.12482
Bigiel F., et al., 2011, ApJ, 730, L13
Bolatto A. D., Wolfire M., Leroy A. K., 2013, ARA&A, 51, 207
Brauer R., Wolf S., Reissl S., Ober F., 2017, A&A, 601, A90
Bubel A.-P., 2015, Master's thesis, Ruprecht-Karls-Universität Heidelberg
Cerviño M., Luridiana V., Pérez E., Vílchez J. M., Valls-Gabaud D., 2003, A&A, 407, 177
Chevance M., et al., 2020a, arXiv e-prints, p. arXiv:2010.13788
Chevance M., et al., 2020b, MNRAS, 493, 2872
Chevance M., et al., 2020c, MNRAS, 494, 5279
Chevance M., Krumholz M. R., McLeod A. F., Ostriker E. C., Rosolowsky E. W., Sternberg A., 2022, arXiv e-prints, p. arXiv:2203.09570
Clark P. C., Glover S. C. O., Klessen R. S., 2012, MNRAS, 420, 745

Clark P. C., Glover S. C. O., Ragan S. E., Duarte-Cabral A., 2019, MNRAS, 486, 4622
Cormier D., et al., 2017, MNRAS, 468, L87
Dabrowski I., 1984, Canadian Journal of Physics, 62, 1639
Dale J. E., Ngoumou J., Ercolano B., Bonnell I. A., 2014, MNRAS, 442, 694
Evans N. J., Kim J.-G., Ostriker E. C., 2022, ApJ, 929, L18
Federrath C., Banerjee R., Seifried D., Clark P. C., Klessen R. S., 2011, in Alves J., Elmegreen B. G., Girart J. M., Trimble V., eds, Proceedings of the International Astronomical Union Vol. 270, Computational Star Formation. pp 425–428 (arXiv:1007.2504), doi:10.1017/S1743921311000755
Girichidis P., Seifried D., Naab T., Peters T., Walch S., Wünsch R., Glover S. C. O., Klessen R. S., 2018, MNRAS, 480, 3511
Glover S. C. O., 2023, in Physics and Chemistry of Star Formation: The Dynamical ISM Across Time and Spatial Scales. p. 242
Glover S. C. O., Clark P. C., 2012, MNRAS, 426, 377
Glover S. C. O., Clark P. C., 2016, MNRAS, 456, 3596
Glover S. C. O., Smith R. J., 2016, MNRAS, 462, 3011
Goldsmith P. F., Velusamy T., Li D., Langer W. D., 2010, ApJ, 715, 1370
Gong M., Ostriker E. C., Wolfire M. G., 2017, ApJ, 843, 38
Gong M., Ostriker E. C., Kim C.-G., Kim J.-G., 2020, ApJ, 903, 142
Graczyk D., et al., 2014, ApJ, 780, 59
Grasha K., et al., 2018, MNRAS, 481, 1016
Grasha K., et al., 2019, MNRAS, 483, 4707
Harris J., Zaritsky D., 2004, AJ, 127, 1531
Hernquist L., 1990, ApJ, 356, 359
Hopkins P. F., Quataert E., Murray N., 2012, MNRAS, 421, 3488
Hu C.-Y., Naab T., Walch S., Glover S. C. O., Clark P. C., 2016, MNRAS, 458, 3528
Hu C.-Y., Naab T., Glover S. C. O., Walch S., Clark P. C., 2017, MNRAS, 471, 2151
Hu C.-Y., Schruba A., Sternberg A., van Dishoeck E. F., 2022, The Astrophysical Journal, 931, 28
Hu C.-Y., Sternberg A., van Dishoeck E. F., 2023, ApJ, 952, 140
Hunter J. D., 2007, Computing in Science Engineering, 9, 90
Hunter G. H., Clark P. C., Glover S. C. O., Klessen R. S., 2023, MNRAS, 519, 4152
Iffrig O., Hennebelle P., 2015, A&A, 576, A95
Inutsuka S.-i., Inoue T., Iwasaki K., Hosokawa T., 2015, A&A, 580, A49
Izquierdo A. F., et al., 2021, MNRAS, 500, 5268
Jeans J. H., 1902, Philosophical Transactions of the Royal Society of London Series A, 199, 1
Kennicutt Jr. R. C., 1998, ApJ, 498, 541
Klessen R. S., Glover S. C. O., 2016, Star Formation in Galaxy Evolution: Connecting Numerical Models to Reality, Saas-Fee Advanced Course, Volume 43.~ISBN 978-3-662-47889-9.~Springer-Verlag Berlin Heidelberg, 2016, p.~85, 43, 85
Kreckel K., et al., 2018, ApJ, 863, L21
Kruijssen J. M. D., Schruba A., Hygate A. P. S., Hu C.-Y., Haydon D. T., Longmore S. N., 2018, MNRAS, 479, 1866
Kruijssen J. M. D., et al., 2019, Nature, 569, 519
Krumholz M. R., McKee C. F., 2005, ApJ, 630, 250
Leitherer C., et al., 1999, ApJS, 123, 3
Leroy A. K., Walter F., Brinks E., Bigiel F., de Blok W. J. G., Madore B., Thornley M. D., 2008, AJ, 136, 2782
Madden S. C., et al., 2020, A&A, 643, A141
Maeder A., 2009, Physics, Formation and Evolution of Rotating Stars. Springer, doi:10.1007/978-3-540-76949-1
Matzner C. D., McKee C. F., 2000, ApJ, 545, 364
McKee C. F., Ostriker E. C., 2007, ARA&A, 45, 565
Omukai K., Hosokawa T., Yoshida N., 2010, ApJ, 722, 1793
Onodera S., et al., 2010, ApJ, 722, L127
Ostriker E. C., McKee C. F., Leroy A. K., 2010, ApJ, 721, 975
Palla F., Salpeter E. E., Stahler S. W., 1983, ApJ, 271, 632
Ramambason L., et al., 2023, arXiv e-prints, p. arXiv:2306.14881
Reissl S., Wolf S., Brauer R., 2016, A&A, 593, A87
Sales L. V., Marinacci F., Springel V., Petkova M., 2014, MNRAS, 439, 2990
Sano H., et al., 2019, ApJ, 881, 85
Schinnerer E., et al., 2019, ApJ, 887, 49
Schöier F. L., van der Tak F. F. S., van Dishoeck E. F., Black J. H., 2005, A&A, 432, 369
Schruba A., Leroy A. K., Walter F., Sandstrom K., Rosolowsky E., 2010, ApJ, 722, 1699
Schruba A., et al., 2012, AJ, 143, 138
Seifried D., Haid S., Walch S., Borchert E. M. A., Bisbas T. G., 2020, MNRAS, 492, 1465
Smith R. J., Glover S. C. O., Clark P. C., Klessen R. S., Springel V., 2014, MNRAS, 441, 1628
Smith R. J., et al., 2020, MNRAS, 492, 1594
Smith M. C., Bryan G. L., Somerville R. S., Hu C.-Y., Teyssier R., Burkhart B., Hernquist L., 2021, MNRAS, 506, 3882
Sormani M. C., Treß R. G., Klessen R. S., Glover S. C. O., 2017, MNRAS, 466, 407
Springel V., 2005, MNRAS, 364, 1105
Springel V., 2010, MNRAS, 401, 791
Stoehr F., Del Prado M., Manning A., Matthews B., Miura R., Plunkett A., Wang K.-S., 2020, in Pizzo R., Deul E. R., Mol J. D., de Plaa J., Verkouter H., eds, Astronomical Society of the Pacific Conference Series Vol. 527, Astronomical Data Analysis Software and Systems XXIX. p. 407
Tokuda K., et al., 2021, ApJ, 922, 171
Tress R. G., Smith R. J., Sormani M. C., Glover S. C. O., Klessen R. S., Mac Low M.-M., Clark P. C., 2020, MNRAS, 492, 2973
Truelove J. K., Klein R. I., McKee C. F., Holliman John H. I., Howell L. H., Greenough J. A., 1997, ApJ, 489, L179
Vázquez-Semadeni E., Colín P., Gómez G. C., Ballesteros-Paredes J., Watson A. W., 2010, ApJ, 715, 1302
Virtanen P., et al., 2020, Nature Methods, 17, 261
Walch S. K., Whitworth A. P., Bisbas T., Wünsch R., Hubber D., 2012, MNRAS, 427, 625
Whitworth D. J., Smith R. J., Tress R., Kay S. T., Glover S. C. O., Sormani M. C., Klessen R. S., 2022, MNRAS, 510, 4146
Whitworth D. J., Smith R. J., Klessen R. S., Mac Low M.-M., Glover S. C. O., Tress R., Pakmor R., Soler J. D., 2023, MNRAS, 520, 89
Wolfire M. G., Hollenbach D., McKee C. F., 2010, ApJ, 716, 1191
van der Walt S., Colbert S. C., Varoquaux G., 2011, Computing in Science Engineering, 13, 22

## APPENDIX A: RESULTS FOR INDIVIDUAL REGIONS

Here we present the full CO emission maps (Figure A1), derived dark gas fractions (Tables A1-A4) and average $X_{\rm CO}$ values (Tables A5-A8) for the individual regions for which we carried out the radiative transfer post-processing. Our naming convention is based on the dendrogram density level where the peak CO density lies, i.e. N14 refers to a structure with a peak CO column density between $10^{14}\,{\rm cm}^{-2}$ and $10^{15}\,{\rm cm}^{-2}$, N13 to a region with a peak CO column density between $10^{13}\,{\rm cm}^{-2}$ and $10^{14}\,{\rm cm}^{-2}$, etc. Different structures sharing the same density level are distinguished by an appended letter (e.g. N13a, N13b, N13c, etc.)

We have computed the CO-to-H$_2$ conversion factor, $X_{\rm CO}$, for each of the selected regions. To calculate $X_{\rm CO}$, we start by taking maps of H$_2$ column density and $W_{\rm CO}$, smoothed at either 14.7″ or 1.5″. We restrict our map of $W_{\rm CO}$ to only those regions with emission above a threshold $W_{\rm CO} = 0.1\,{\rm K\,km\,s}^{-1}$ and then produce a map of $X_{\rm CO}$ for each region by dividing the H$_2$ column density map by the $W_{\rm CO}$ map. This gives us a smoothed map of $X_{\rm CO}$ for each region. We then average the values in this smoothed map to arrive at a single representative value. These values are reported in Tables A5–A8. Our restriction of this analysis to regions with $W_{\rm CO} > 0.1\,{\rm K\,km\,s^{-1}}$ is motivated by the fact that fainter re-gions are unlikely to be detectable. We report no value for fully dark regions in which there is no pixel with CO emission above the threshold; note that this is the majority of regions in simulation Z.10 G.10 and Z.01 G.10.

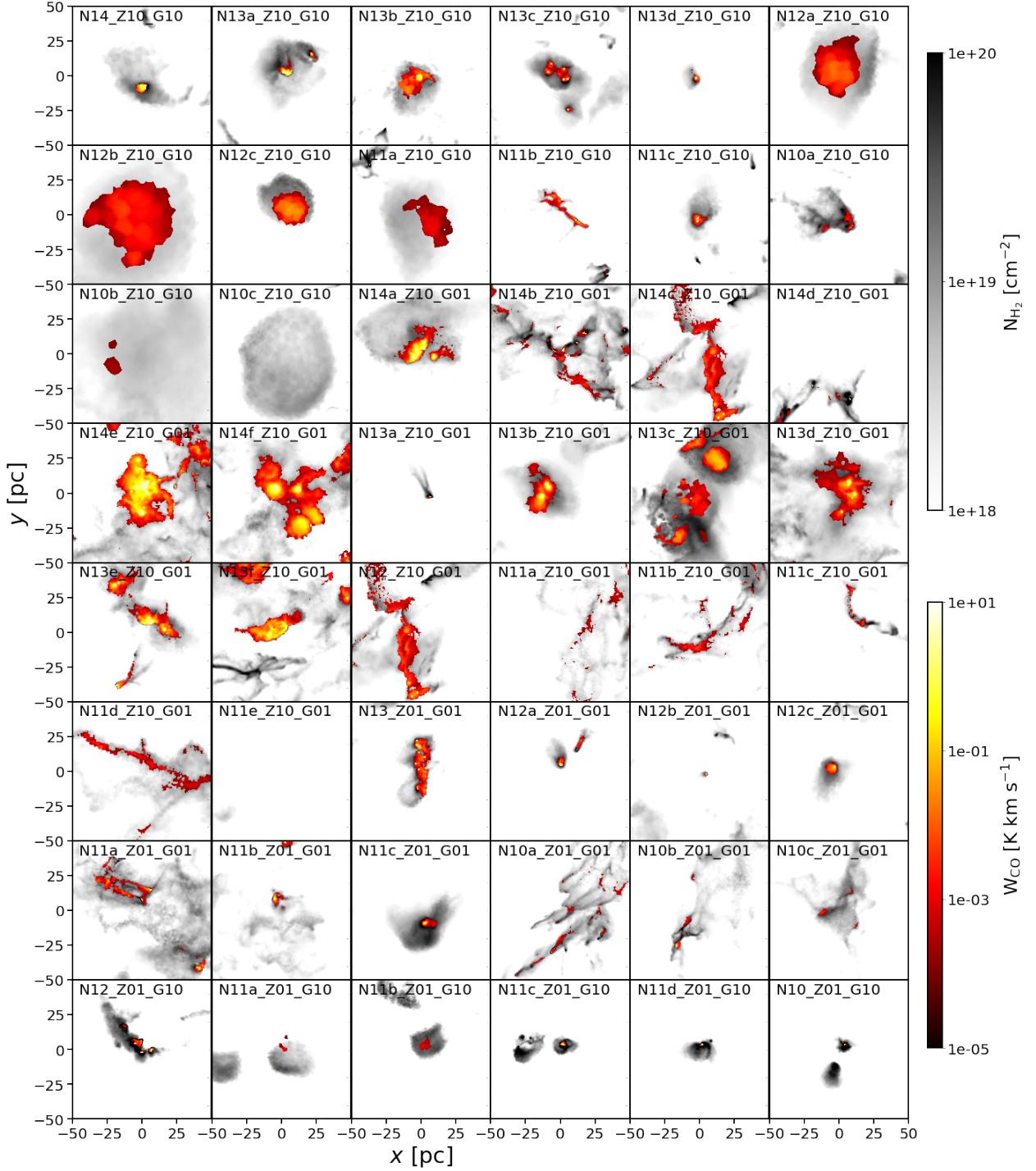

**Figure A1.** Maps for all of the regions used in the analysis showing $W_{\rm CO}$ wherever it exceeds $10^{-5}$ K km s$^{-1}$, traced on top of H$_2$ surface density. We can see a variety of shapes, sizes and density distributions across the regions, ranging from diffuse "blobs" to dense core-like structures and filaments. Note that the $W_{\rm CO}$ maps shown here are the direct output from POLARIS, and so the irregular shapes of the Voronoi cells are sometimes apparent.

For the regions that have at least some detectable CO emission, we see that there is substantial variability in their values of $X_{\rm CO}$, similar to the result recently reported by Hu et al.

| Region | $f_{\rm DG}$ 14.7″ | $f_{\rm DG}$ 1.5″ |
| --- | --- | --- |
| N14 Z.10 G.10 | 1.00 | 1.00 |
| N13a Z.10 G.10 | 1.00 | 0.97 |
| N13b Z.10 G.10 | 1.00 | 1.00 |
| N13c Z.10 G.10 | 1.00 | 1.00 |
| N13d Z.10 G.10 | 1.00 | 1.00 |
| N12a Z.10 G.10 | 1.00 | 1.00 |
| N12b Z.10 G.10 | 1.00 | 1.00 |
| N12c Z.10 G.10 | 1.00 | 1.00 |
| N11a Z.10 G.10 | 1.00 | 0.86 |
| N11b Z.10 G.10 | 1.00 | 1.00 |
| N11c Z.10 G.10 | 1.00 | 1.00 |
| N10a Z.10 G.10 | 1.00 | 1.00 |
| N10b Z.10 G.10 | 1.00 | 1.00 |
| N10c Z.10 G.10 | 1.00 | 1.00 |
| Average Z.10 G.10 | 1.00 | 1.00 |

**Table A1.** Dark gas fractions ($f_{\rm DG}$) for Z.10 G.10 across all regions in the simulation. The average shown is the mass-weighted average of the regions (the black line in Figure 6) and not the average of the individual $f_{\rm DG}$ values.

| Region | $f_{\rm DG}$ 14.7″ | $f_{\rm DG}$ 1.5″ |
| --- | --- | --- |
| N14a Z.10 G.01 | 0.37 | 0.42 |
| N14b Z.10 G.01 | 0.57 | 0.55 |
| N14c Z.10 G.01 | 0.83 | 0.77 |
| N14d Z.10 G.01 | 0.34 | 0.41 |
| N14e Z.10 G.01 | 0.35 | 0.46 |
| N14f Z.10 G.01 | 0.59 | 0.65 |
| N13a Z.10 G.01 | 0.06 | 0.10 |
| N13b Z.10 G.01 | 0.36 | 0.56 |
| N13c Z.10 G.01 | 1.00 | 1.00 |
| N13d Z.10 G.01 | 0.49 | 0.62 |
| N13e Z.10 G.01 | 0.35 | 0.47 |
| N13f Z.10 G.01 | 0.50 | 0.54 |
| N12 Z.10 G.01 | 0.81 | 0.82 |
| N11a Z.10 G.01 | 1.00 | 1.00 |
| N11b Z.10 G.01 | 1.00 | 1.00 |
| N11c Z.10 G.01 | 1.00 | 1.00 |
| N11d Z.10 G.01 | 1.00 | 1.00 |
| N11e Z.10 G.01 | 1.00 | 1.00 |
| Average Z.10 G.01 | 0.49 | 0.13 |

**Table A2.** As Table A1, but for simulation Z.10 G.01.

| Region | $f_{\rm DG}$ 14.7″ | $f_{\rm DG}$ 1.5″ |
| --- | --- | --- |
| N13 Z.01 G.01 | 1.00 | 1.00 |
| N12a Z.01 G.01 | 1.00 | 1.00 |
| N12b Z.01 G.01 | 1.00 | 1.00 |
| N12c Z.01 G.01 | 1.00 | 1.00 |
| N11a Z.01 G.01 | 1.00 | 1.00 |
| N11b Z.01 G.01 | 1.00 | 1.00 |
| N11c Z.01 G.01 | 1.00 | 1.00 |
| N10a Z.01 G.01 | 1.00 | 1.00 |
| N10b Z.01 G.01 | 0.48 | 0.57 |
| N10c Z.01 G.01 | 1.00 | 1.00 |
| Average Z.01 G.01 | 1.00 | 0.89 |

**Table A3.** As Table A1, but for simulation Z.01 G.01.

| Region | $f_{\rm DG}$ 14.7″ | $f_{\rm DG}$ 1.5″ |
| --- | --- | --- |
| N12 Z.01 G.10 | 0.60 | 0.43 |
| N11a Z.01 G.10 | 1.00 | 1.00 |
| N11b Z.01 G.10 | 1.00 | 1.00 |
| N11c Z.01 G.10 | 0.62 | 0.65 |
| N11d Z.01 G.10 | 0.68 | 0.49 |
| N10 Z.01 G.10 | 1.00 | 0.61 |
| Average Z.01 G.10 | 1.00 | 0.50 |

**Table A4.** As Table A1, but for simulation Z.01 G.10.

| Region | $X_{\rm CO}$ 14.7″ | $X_{\rm CO}$ 1.5″ |
| --- | --- | --- |
| N14 Z.10 G.10 | - | - |
| N13a Z.10 G.10 | - | 4.72 $\times 10^{22}$ |
| N13b Z.10 G.10 | - | 2.94 $\times 10^{19}$ |
| N13c Z.10 G.10 | - | 5.06 $\times 10^{21}$ |
| N13d Z.10 G.10 | - | - |
| N12a Z.10 G.10 | - | - |
| N12b Z.10 G.10 | - | - |
| N12c Z.10 G.10 | - | - |
| N11a Z.10 G.10 | - | 9.69 $\times 10^{22}$ |
| N11b Z.10 G.10 | - | - |
| N11c Z.10 G.10 | - | - |
| N10a Z.10 G.10 | - | - |
| N10b Z.10 G.10 | - | - |
| N10c Z.10 G.10 | - | - |

**Table A5.** Average CO-to-H$_2$ conversion factor ($X_{\rm CO}$, units are cm$^{-2}$ (K km s$^{-1}$)$^{-1}$) for the regions selected from simulation Z.10 G.10. The values of $X_{\rm CO}$ are calculated using maps of $W_{\rm CO}$ and $N_{\rm H_2}$ smoothed on the indicated scale. The final reported value is then an average of the values computed for the region. We do not report a value for regions that have $W_{\rm CO} < 0.1$ K km s$^{-1}$, since these are unlikely to be detectable in CO.

(2022). In our runs with $Z = 0.1\,Z_\odot$, we find typical value of $X_{\rm CO}$ ranging from a few times smaller to a few tens of times larger than the standard Galactic value. This broadly agrees with the result of the fiducial time-dependent model in Hu et al. (2022) for a galaxy with the same metallicity, but is considerably smaller than most observational determinations at the same metallicity (see e.g. Schruba et al. 2012). However, this is likely due to the fact that we are reporting values purely for CO-bright regions whereas the observations are typically probing scales encompassing both CO-bright and CO-dark H$_2$. For our $Z = 0.01\,Z_\odot$ models, it is harder to draw firm conclusions, given the small number of structures that we find with detectable CO and the paucity of observational detections of molecular gas in such environments.

## APPENDIX B: REGION SIZE STUDY

In the main text, we selected a $(3\,{\rm pc})^3$ sub-region around each sink for further analysis. This choice was motivated by the desire to sample not just the dense, star-forming clumps but also the more diffuse gas surrounding them, while also keeping the regions small enough to minimize confusion between different star-forming clumps. To explore how sensitive our results are to this choice, we have repeated part of our analysis for larger $(5\,{\rm pc})^3$ and $(10\,{\rm pc})^3$ cubical boxes around each sink.

| Region | $X_{\rm CO}$ | |
|---|---|---|
| | $14.7''$ | $1.5''$ |
| N14a Z.10 G.01 | $3.60 \times 10^{20}$ | $2.62 \times 10^{20}$ |
| N14b Z.10 G.01 | $2.67 \times 10^{21}$ | $3.02 \times 10^{21}$ |
| N14c Z.10 G.01 | $1.89 \times 10^{21}$ | $2.35 \times 10^{21}$ |
| N14d Z.10 G.01 | $1.93 \times 10^{21}$ | $1.55 \times 10^{21}$ |
| N14e Z.10 G.01 | $6.99 \times 10^{20}$ | $5.52 \times 10^{20}$ |
| N14f Z.10 G.01 | $3.14 \times 10^{20}$ | $3.09 \times 10^{20}$ |
| N13a Z.10 G.01 | $5.70 \times 10^{21}$ | $4.50 \times 10^{21}$ |
| N13b Z.10 G.01 | $7.69 \times 10^{20}$ | $8.91 \times 10^{20}$ |
| N13c Z.10 G.01 | - | $3.88 \times 10^{20}$ |
| N13d Z.10 G.01 | $1.65 \times 10^{21}$ | $1.42 \times 10^{21}$ |
| N13e Z.10 G.01 | $6.76 \times 10^{20}$ | $4.29 \times 10^{20}$ |
| N13f Z.10 G.01 | $1.27 \times 10^{21}$ | $9.07 \times 10^{20}$ |
| N12 Z.10 G.01 | $1.88 \times 10^{21}$ | $2.16 \times 10^{21}$ |
| N11a Z.10 G.01 | - | - |
| N11b Z.10 G.01 | - | - |
| N11c Z.10 G.01 | - | - |
| N11d Z.10 G.01 | - | - |
| N11e Z.10 G.01 | - | - |

**Table A6.** As Table A5, but for simulation Z.10 G.01.

| Region | $X_{\rm CO}$ | |
|---|---|---|
| | $14.7''$ | $1.5''$ |
| N13 Z.01 G.01 | - | - |
| N12a Z.01 G.01 | - | - |
| N12b Z.01 G.01 | - | - |
| N12c Z.01 G.01 | - | - |
| N11a Z.01 G.01 | - | - |
| N11b Z.01 G.01 | - | - |
| N11c Z.01 G.01 | - | - |
| N10a Z.01 G.01 | - | - |
| N10b Z.01 G.01 | $3.54 \times 10^{21}$ | $3.42 \times 10^{21}$ |
| N10c Z.01 G.01 | - | - |

**Table A7.** As Table A5, but for simulation Z.01 G.01.

| Region | $X_{\rm CO}$ | |
|---|---|---|
| | $14.7''$ | $1.5''$ |
| N12 Z.01 G.10 | $4.27 \times 10^{21}$ | $6.30 \times 10^{21}$ |
| N11a Z.01 G.10 | - | - |
| N11b Z.01 G.10 | - | - |
| N11c Z.01 G.10 | $3.93 \times 10^{21}$ | $3.49 \times 10^{21}$ |
| N11d Z.01 G.10 | $4.01 \times 10^{21}$ | $4.06 \times 10^{21}$ |
| N10 Z.01 G.10 | - | $5.27 \times 10^{21}$ |

**Table A8.** As Table A5, but for simulation Z.01 G.10.

If we compare Figures B1 and B2 to the main results in Figure 7 we see that the mass-weighted $H_2$ densities (the bottom row in each plot) change very little. This is to be expected, since the dense clumps containing most of the $H_2$ mass are much smaller than the size of the box. Enlarging the box adds little additional low-density $H_2$ and hence has little impact on the mass-weighted $H_2$ number density. For the same reason, however, enlarging the box does significantly affect the volume-weighted $H_2$ number density, which drops from $\sim 100\,{\rm cm}^{-3}$ with the $(3\,{\rm pc})^3$ box to $\sim 10\,{\rm cm}^{-3}$ with the $(10\,{\rm pc})^3$ box, reflecting the fact that we are averaging the density of a moderately larger amount of $H_2$ over a much larger volume.

| Sink Age (Myr) | Total gas ($\rm M_\odot$) | Total $H_2$ ($\rm M_\odot$) | $T \leq 300\,{\rm K}$ ($\rm M_\odot$) |
|---|---|---|---|
| $(5\,{\rm pc})^3$ | | | |
| 1 | 104828 | 62755 | 100591 |
| 2 | 64991 | 39308 | 60581 |
| 3 | 42886 | 28030 | 39526 |
| 4 | 35264 | 21373 | 29600 |
| 5 | 35621 | 26857 | 32328 |
| 6 | 19196 | 13395 | 17607 |
| 7 | 19476 | 13222 | 17595 |
| 8 | 20758 | 15394 | 18773 |
| $(10\,{\rm pc})^3$ | | | |
| 1 | 165468 | 81892 | 154035 |
| 2 | 101199 | 51623 | 89273 |
| 3 | 77770 | 41639 | 65765 |
| 4 | 76239 | 37791 | 55179 |
| 5 | 62266 | 39363 | 51449 |
| 6 | 33919 | 19785 | 28421 |
| 7 | 36537 | 21221 | 29869 |
| 8 | 32704 | 19980 | 27657 |

**Table B1.** The total gas mass, molecular gas mass and mass of gas below 300 K in 5 pc and 10 pc cubes surrounding sinks of different ages in all models. As the size of the cube increases, the fraction of the total mass that is molecular decreases, but the gas remains mostly cold.

Looking at the amount of gas close to the sinks that is cold and/or molecular tells us a similar story. If we compare the results in Table B1 for the $(5\,{\rm pc})^3$ and $(10\,{\rm pc})^3$ boxes with the results shown in Table 6 for our fiducial $(3\,{\rm pc})^3$ box, we see that the fraction of the gas around the sinks that is molecular decreases as the volume increases, consistent with the idea that much of the additional volume is diffuse atomic gas. However, we also see that even with the large boxes, the gas around the sinks is mostly cold.

## APPENDIX C: WEIGHTING CHOICE

We show both the mass and volume weighted number densities in the main text as when we compare between mass and volume weighted densities we see a discrepancy between the two. AREPO can hyper-refine beyond the sink creation density creating an abundance of very small, very high density cells. It also allows the cells to vary in mass by a factor of 2 from the base cell mass limit of $1\,{\rm M}_\odot$. To see which weighting method is more viable we examine the properties and distribution of the cells.

In the main text, we showed both the mass-weighted average and the volume weighted-average of the $H_2$ number density in the gas near the sinks, and found that there was a large difference between these two values. Here, we look in a little more detail at the reasons for this difference.

In Figure C1 we show the number of cells as a function of number density for the regions around the sinks for three different sink ages: 1 Myr, 3 Myr and 8 Myr. We see a clear peak at a number density of $10^5 \, \text{cm}^{-3}$, the sink creation density. The sharp fall-off above this density is a consequence of the sink formation algorithm: in general, cells can reside above this threshold for only a short time before their gas is either converted to form a sink or accreted by an existing sink. The sharp drop-off in the number of cells at lower densities is a consequence of the Jeans refinement criterion in the simulations: as the density increases, the Jeans length drops and so the code refines the cells to ensure that the Jeans length continues to be resolved. As a result, the mass associated with each cell drops as we move to higher densities.

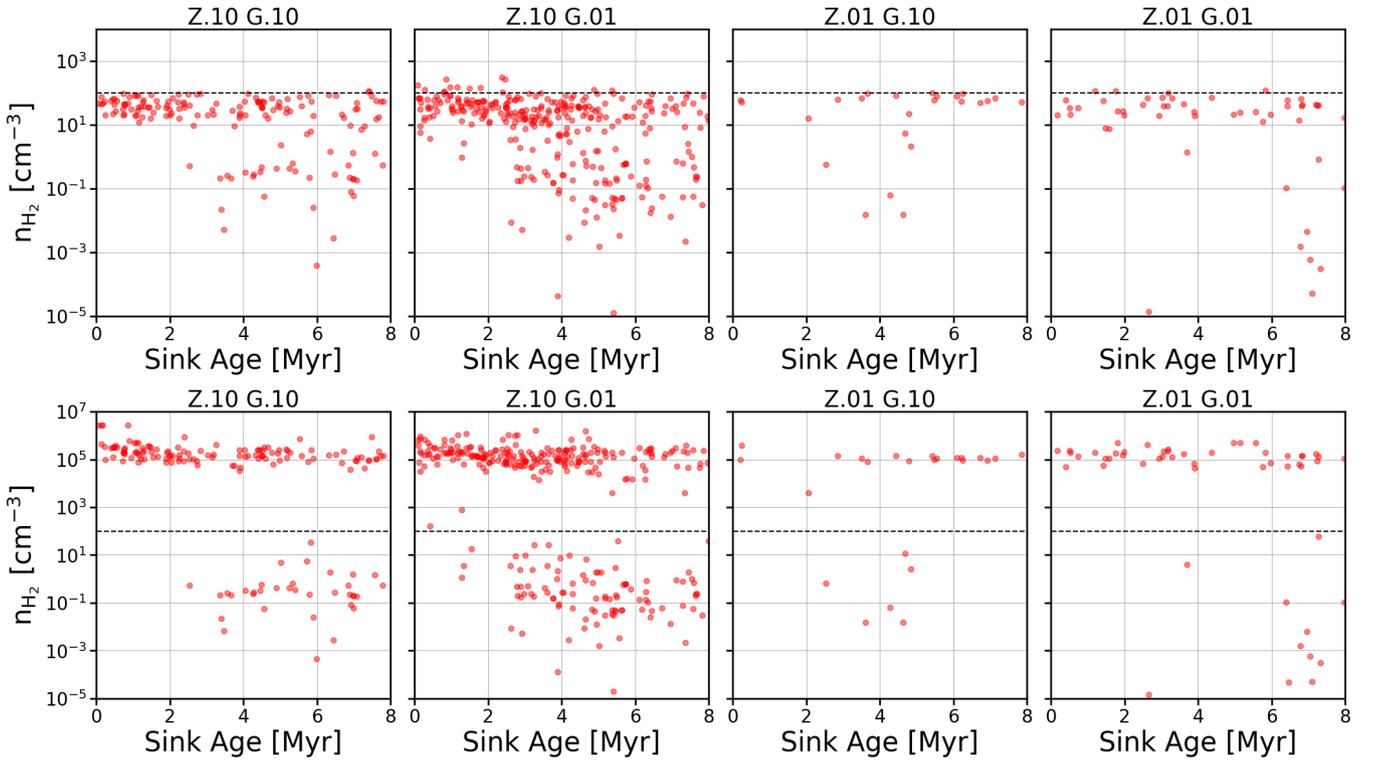

**Figure B1.** Top: The volume-weighted number density of $H_2$ in a $(5\,\text{pc})^3$ cube centred on a sink plotted in relation to the age of the sink. Bottom: The mass-weighted number density of $H_2$ in a $(5\,\text{pc})^3$ cube centred on a sink plotted in relation to the age of the sink. The dashed line represents the turn over to dense gas at a number density of $100\,\text{cm}^{-3}$.

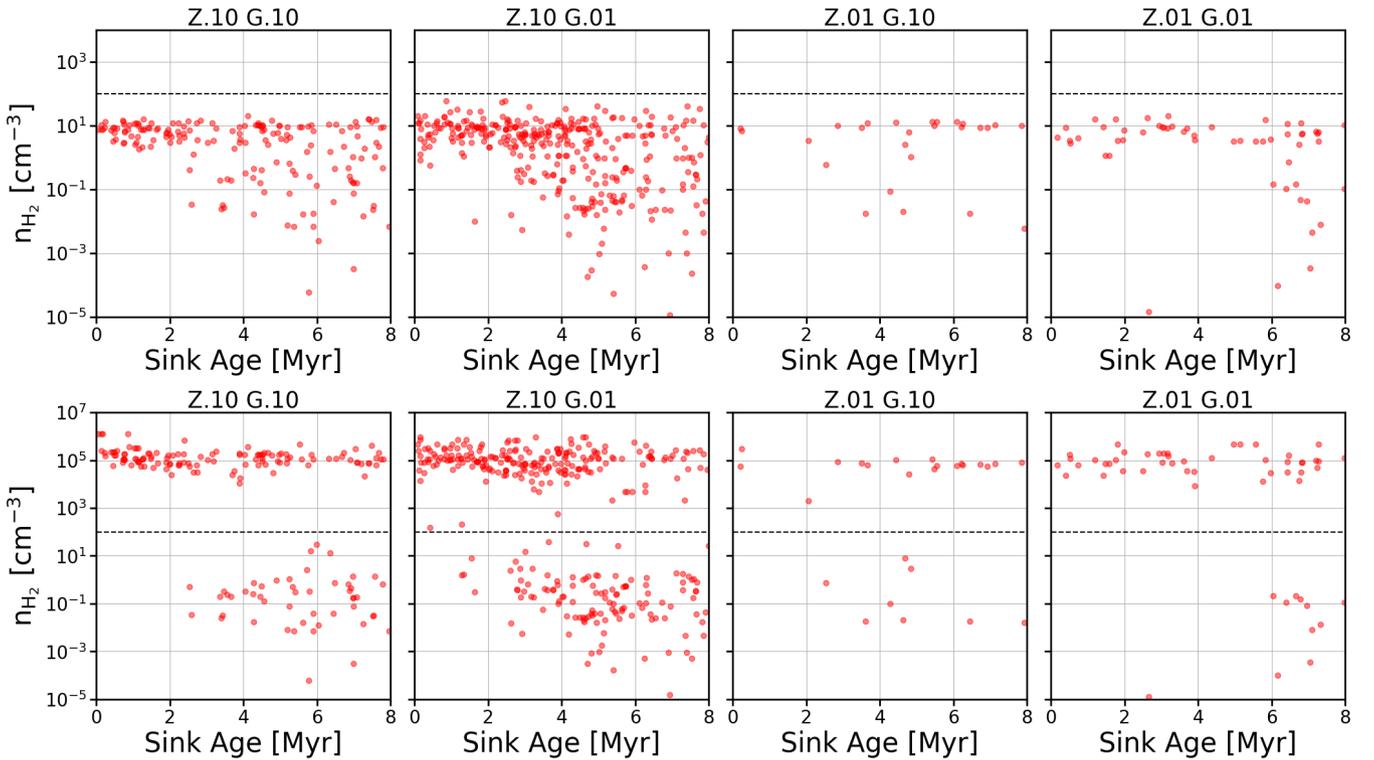

**Figure B2.** As Figure B1, but for a $(10\,\text{pc})^3$ cube centred on each sink.

The impact of the refinement scheme becomes clear if we compare these plots with plots showing the mass distribution as a function of density for the same regions (left-hand panels in Figure C3). The peak at a density of $10^5 \, \text{cm}^{-3}$ remains, but is less prominent, since these cells all have small masses. A second peak is also visible in the distribution, located at a density of $\sim 100$–$1000 \, \text{cm}^{-3}$. This corresponds to the characteristic range of densities in the gas clouds that host the dense, star-forming clumps; as the volume-weighted distribution demonstrates (right-hand panels in Figure C3), this gas fills the bulk of the volume around the sinks, with the dense clumps taking up only a small fraction of the total volume. We see therefore that the mass-weighted average gives us a better view of the conditions in the clumps, whereas the volume-weighted average better represents the conditions in the clouds on larger scales. That said, neither weighting method is a completely true representation of the properties of the gas around the sinks, and so for this reason we have chosen to present both in this work.

This paper has been typeset from a TeX/LaTeX file prepared by the author.

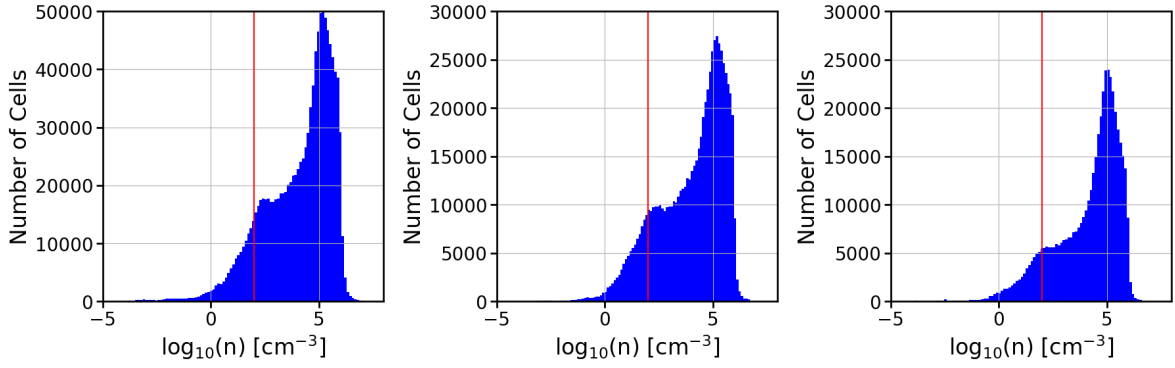

**Figure C1.** The number of cells around sinks that are aged 1 Myr (left), 3 Myr (middle) and 8 Myr (right) as a function of density. The red line shows where n = 100 cm$^{-3}$.

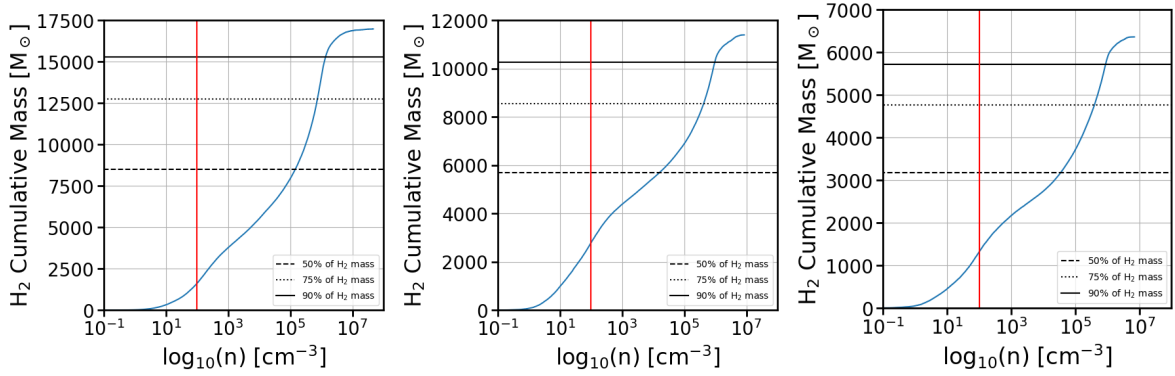

**Figure C2.** The cumulative mass of H$_2$ around sinks that are aged 1 Myr (left), 3 Myr (middle) and 8 Myr (right) as a function of density. The red line shows where n = 100 cm$^{-3}$, with the black lines showing where 50% of the H$_2$ mass, 75% of the H$_2$ mass and 90% of the H$_2$ mass lie.

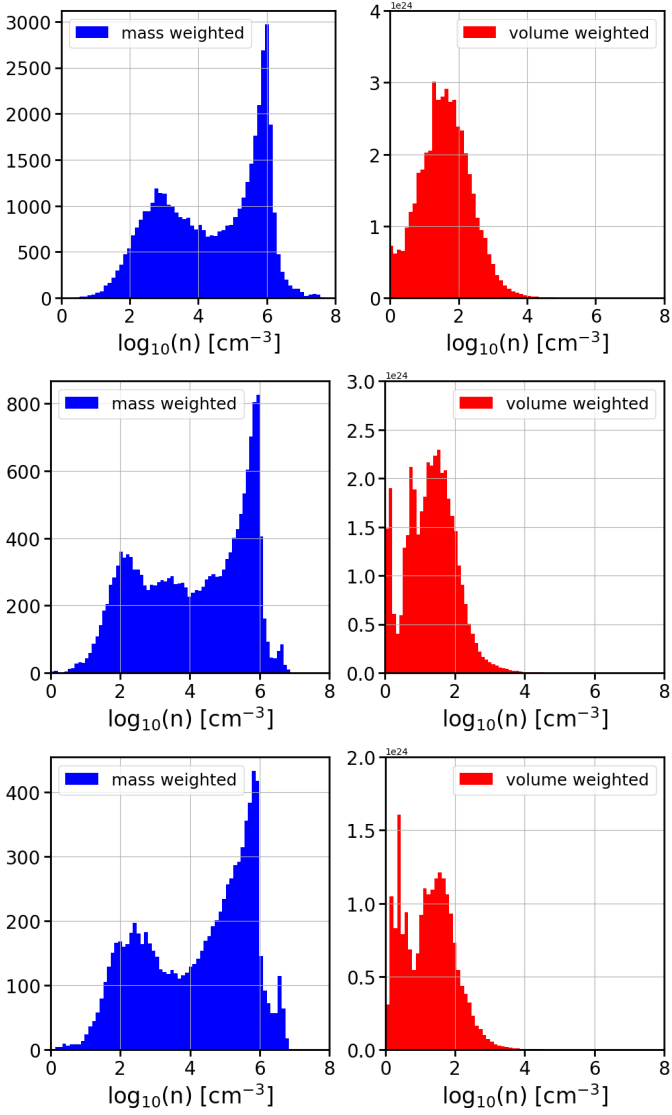

**Figure C3.** Mass-weighted and volume-weighted histograms for the cells around sinks at ages of 1 Myr (top), 3 Myr (middle) and 8 Myr (bottom).